\documentclass[twocolumn,amsmath,amssymb,nofootinbib]{revtex4}
\usepackage{graphicx}
\usepackage{dcolumn}
\usepackage{bm}
\usepackage{amsmath,amssymb}
\usepackage{color}
\usepackage{longtable}
\usepackage{ulem}

\newcommand\Ocal{\mathcal{O}}
\newcommand\Vcal{\mathcal{V}}

\newcommand\QQbar{Q\overline{Q}}
\newcommand\mQ{m_{Q}}
\newcommand\SdotS{{\mathbf S}_Q\cdot{\mathbf S}_{\overline{Q}}}

\newcommand\SQ{{\mathbf S}_Q}
\newcommand\SQbar{{\mathbf S}_{\overline{Q}}}
\newcommand\tsrc{t_{\rm s}}

\usepackage[colorlinks=true, linkcolor=blue, filecolor=blue, urlcolor=blue, citecolor=blue, pdftex=true, plainpages=false]{hyperref}

\begin{document}
\title{ Heavy quarkonium potential from Bethe-Salpeter wave function on the lattice }


\author{Taichi Kawanai${}^{1,2}$} \email{kawanai@nt.phys.s.u-tokyo.ac.jp}
\author{Shoichi Sasaki${}^{1,3}$} \email{ssasaki@nucl.phys.tohoku.ac.jp}
\affiliation{${}^{1}$Theoretical Research Division, Nishina Center, RIKEN, Wako 351-0198, Japan}
\affiliation{${}^{2}$Department of Physics, Brookhaven National Laboratory Upton, NY 11973-5000, USA} 
\affiliation{${}^{3}$Department of Physics, Tohoku University, Sendai 980-8578, Japan}

\date{\today}

\begin{abstract}
We propose a novel method for the determination of the
interquark potential together with quark ``kinetic mass'' $\mQ$ 
from the equal-time $\QQbar$ Bethe-Salpeter~(BS) amplitude in lattice QCD.
Our approach allows
us to calculate spin-dependent $\QQbar$ potentials, {\it e.g.} the spin-spin potential, 
as well. 
In order to investigate several systematic uncertainties
on such $\QQbar$ potentials,
we carry out lattice QCD simulations using quenched gauge configurations generated
with the single plaquette gauge action with three different lattice spacings,
$a \approx  0.093, 0.068$ and  $0.047$~fm,
and two different physical volumes, $L \approx 2.2$ and $3.0 $~fm.
For heavy quarks, we employ the relativistic heavy quark~(RHQ) action which 
can control large discretization errors introduced by large quark mass $\mQ$.
The spin-independent central $\QQbar$ potential for the charmonium system yields the
``Coulomb plus linear'' behavior with good scaling and small volume dependence.
We explore the quark mass dependence over the wide mass range 
from the charm to beyond the bottom region, and then demonstrate that the spin-independent central 
$\QQbar$ potential in the $\mQ \to \infty$ limit is fairly consistent with the static $\QQbar$ potential  
obtained from Wilson loops.
The spin-spin potential at finite quark mass provides a repulsive interaction with a finite range, 
which becomes narrower as the quark mass increases.
We also discuss the applicability of the $1/\mQ$ expansion approach for the spin-spin potential.
\end{abstract}

\maketitle


\section{\label{intro} Introduction }
The dynamics of heavy quarks
having much larger masses than the QCD scale, $\Lambda_{\rm QCD}$,
can be analyzed within the formalism of nonrelativistic quantum mechanics.
In quark potential models, physical quantities such as mass spectra and
decay rates of heavy quarkonium states are indeed calculated by solving 
the Schr\"odinger equation with heavy ``constituent quark mass''~\cite{Close:1979bt}.
The so-called Cornell potential is often adopted 
as an interquark potential between a heavy quark ($Q$) and an anti-quark 
($\overline{Q}$)~\cite{Eichten:1974af}. 

The Cornell potential is consisted of a Coulomb part and a linear part as  
\begin{equation}
V(r) = -\frac{4}{3}\frac{\alpha_s}{r} + \sigma r,
\end{equation}
where $\alpha_s$ is the strong coupling constant
and $\sigma$ denotes the string tension~\cite{Eichten:1974af}.
The first term is generated by perturbative one-gluon exchange, 
while the linearly rising potential describes the phenomenology of confining quark interactions.
Indeed confining nature of QCD is a key ingredient for 
understanding heavy quarkonium physics~\cite{Eichten:1974af,Eichten:1978tg, Eichten:1979ms, Richardson:1978bt,Barnes:2005pb}.

Spin-dependent potentials~(spin-spin, tensor and spin-orbit terms)
appear as relativistic corrections to the spin-independent central potential
in powers of the inverse of the heavy quark mass
$\mQ$~\cite{Eichten:1980mw,Godfrey:1985xj}.
In potential models, their functional forms are basically determined by 
perturbative one-gluon exchange as
the Fermi-Breit type potential~\cite{Eichten:1980mw,Godfrey:1985xj}. 
For heavy quarkonia ($\mQ=m_{\overline{Q}}$), the spin-dependent potentials are give by
\begin{multline}
 V_{\rm spin-dep} = \frac{1}{m_Q^2}\Big[ \frac{32\pi\alpha_s}{9} \delta(r) \SQ \cdot \SQbar \\
+ \frac{4\alpha_s}{r^3}\left(\frac{(\SQ\cdot{\bf r})(\SQbar\cdot{\bf r})}{r^2}-\frac{\SQ\cdot \SQbar}{3}\right)
+ \left( \frac{2\alpha_s}{r^3}- \frac{\sigma}{2r}\right) {\bf L} \cdot {\bf S} \Big],
\end{multline}
where ${\bf S}=\SQ + \SQbar$.
Although there are many successes in the conventional charmonium spectrum, 
many of the newly discovered charmonium-like mesons, 
named as ``$XYZ$'' mesons,
could not be simply explained by quark potential models~\cite{Godfrey:2008nc}. 
However, the phenomenological spin-dependent potentials
determined by perturbative method would have validity only at short distances and also
in the vicinity of the heavy quark mass limit.
We thus consider that properties of higher-mass charmonium states predicated 
in quark potential models may suffer from large uncertainties.

In order to make more accurate theoretical predictions
in quark potential models, the reliable interquark potential directly derived
from first principles QCD is highly desired.
One of the major successes of lattice QCD is to qualitatively justify 
the Coulomb plus linear potential by the static heavy quark potential
obtained from Wilson loops~\cite{Bali:2000gf}.
Indeed, the $\QQbar$ potential between infinitely heavy quark
and antiquark has been precisely determined 
by lattice QCD in the past few decades~\cite{{Bali:1992ab}, {Schilling:1993bk},
 {Bali:1992ru}, {Booth:1992bm}, {Bali:2000gf}, {Glassner:1996xi}}.

The relativistic corrections to the static potential are
classified in powers of $1/\mQ$ within a framework called potential
nonrelativistic QCD (pNRQCD)~\cite{Brambilla:2004jw}.
The lattice determination of the spin-dependent terms 
has been carried out within the quenched approximation
in the 1980s~\cite{deForcrand:1985zc, Michael:1985wf, Michael:1985rh, Huntley:1986de, Campostrini:1986ki,Campostrini:1987hu}
and the 1990s~\cite{Bali:1996cj, Bali:1997am},
and it has been extended to dynamical QCD simulations~\cite{Koike:1989jf,Born:1993cp}. 
However, these earlier studies did not enable us to determine the functional forms
of the spin-dependent terms due to large statistical errors.

Recently, corrections of the leading and next-to-leading order in the $1/\mQ$ expansion
to the static $\QQbar$ potential have been successfully calculated in {\it quenched} lattice QCD
with high accuracy by using  the multilevel algorithm~\cite{Koma:2006fw,Koma:2010zz}.
However calculation of the realistic {\it charmonium} potential in lattice QCD within Wilson loop formalism
is still rather difficult.
The inverse of the charm quark mass would be obviously 
far outside the validity region of the $1/\mQ$ expansion.
Furthermore, a spin-spin potential determined at 
${\cal O}(1/\mQ^2)$~\cite{Koma:2006fw,Koma:2010zz},
which provides an attractive interaction for the higher spin states,
yields wrong mass ordering among hyperfine multiplets. 
The higher order corrections beyond the next-to-leading order are
required to correctly describe the conventional heavy quarkonium spectrum
{\it even for the bottom quarks} with the  $\QQbar$ 
potentials obtained from this approach.
In addition, practically, the multilevel algorithm employed
in Refs.~\cite{Koma:2006fw,Koma:2010zz} is not easy
to be implemented in dynamical lattice QCD simulations.

Under these circumstances, we propose a novel method to determine 
the interquark potential using lattice QCD in this paper.
The interquark potential is defined by the equal-time and Coulomb gauge Bethe-Salpeter (BS)
amplitude through an effective Schr\"odinger equation.
This is a variant of the mehod originally applied for the hadron-hadron interaction~\cite{Ishii:2006ec,Aoki:2009ji},
and enables us to determine both spin-independent and spin-dependent
interquark potentials, {\it at heavy, but finite quark mass}.
These potentials implicitly  account for all orders of
$1/\mQ$ corrections~\cite{Kawanai:2011xb}. 
Furthermore, there is no restriction to dynamical calculation within this method.

This paper is organized as follows.
In Sec.~\ref{formalism}, we briefly review the methodology utilized in this paper
to calculate the spin-independent and spin-dependent $\QQbar$ potentials 
with the finite quark mass in lattice QCD simulations.
Sec.~\ref{setup} contains details of our Monte Carlo simulations 
and some basic results.
In Sec.~\ref{main}, we show numerical results of the quark kinetic mass $\mQ$,
the spin-independent central and spin-spin potentials.
The spin-independent part of the resulting $\QQbar$ potential exhibits 
a good scaling behavior and small volume dependence.
We also discuss several systematic uncertainties on the interquark potentials 
obtained from the BS amplitude.
In Sec.~\ref{limit}, 
in order to demonstrate the feasibility of the new approach,
we show that the interquark potential calculated by the BS amplitude 
smoothly approaches the  static $\QQbar$ potential 
from Wilson loops in the infinitely heavy quark limit.
We also discuss an issue on the spin-spin potential in the conventional $1/\mQ$ expansion approach.
In Sec.~\ref{summary},
we summarize and discuss all results and future perspectives.

%
%
\section{\label{formalism} Formalism}
In this section, we will briefly review the new method utilized here to
calculate the interquark potential {\it at finite quark mass}.
Our proposed method for a new determination of the interquark potential in lattice QCD
is based on the same idea originally applied for the nuclear force~\cite{Ishii:2006ec,Aoki:2009ji},
where the hadron-hadron potential is defined through the equal-time BS
amplitude~\cite{Ishii:2006ec,Aoki:2009ji, Nemura:2008sp,Ikeda:2010zz,Kawanai:2010ev,Doi:2011gq,Aoki:2012tk,Murano:2011nz,Aoki:2011gt,HALQCD:2012aa}.
Here we call this method as {\it BS amplitude method}.
The quark kinetic mass $\mQ$, which is a key ingredient in the BS amplitude method applied to the $\QQbar$ system,
is simultaneously determined through the large-distance behavior in the spin-dependent
part of the interquark potential with the help of the measured hyperfine splitting energy of
$1S$ states in heavy quarkonia~\cite{Kawanai:2011xb,Kawanai:2011jt}.

\subsection{Equal-time $\QQbar$ BS wave function}
A gauge-invariant definition of the equal-time $\QQbar$
BS amplitude for quarkonium states is given by
\begin{equation}
  \phi_\Gamma({\bf r})= \sum_{{\bf x}}\langle 0| \overline{Q}
  ({\bf x})\Gamma {\cal M}({\bf x}, {\bf x}+{\bf r})Q({\bf x}+{\bf r})|
  \QQbar;J^{PC}\rangle \label{eq_phi_gauge_inv},
\end{equation}
where ${\bf r}$ is the relative coordinate of two quarks at a certain 
time slice, and $\Gamma$ is any of the 16 Dirac $\gamma$ 
matrices~\cite{{Velikson:1984qw},{Gupta:1993vp}}.
A summation over spatial coordinates ${\bf x}$ projects 
on a state with zero total momentum.
${\cal M}$ is a path-ordered product of gauge links. 
The ${\bf r}$-dependent amplitude, $\phi_\Gamma({\bf r})$, is here called BS 
wave function. In the Coulomb or Landau gauge, the BS wave function 
can be simply evaluated with ${\cal M}=1$. Hereafter, we consider 
the {\it Coulomb gauge} BS wave function~\cite{{Gupta:1993vp}}.

 \begin{figure}
   \centering
   \includegraphics[width=.4\textwidth]{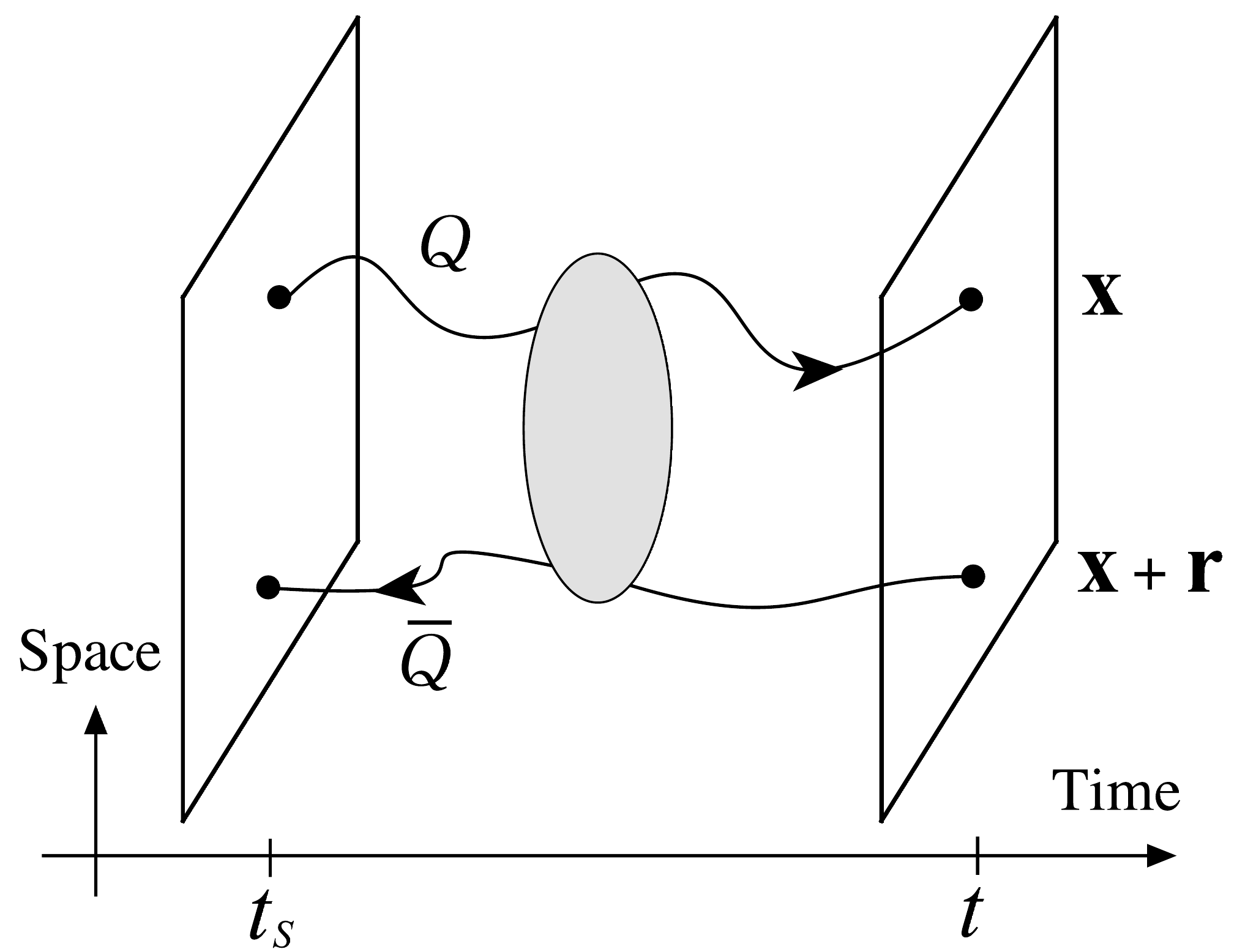}
  \caption{A schematic illustration of the four-point correlation function for the $\QQbar$ system.
    Solid lines indicate quark propagators of a heavy quark and a heavy antiquark,
    located at ${\bf x}$ and ${\bf x}+ {\bf r}$ at sink, respectively.    
    At the source~($t_S$) and sink~($t$), the gauge field configurations are fixed 
    to the Coulomb gauge.}
  \label{fourpoint}
 \end{figure}
The Coulomb gauge BS wave function can be extracted from 
the following four-point correlation function
that is also schematically depicted in Fig.~\ref{fourpoint}:
\begin{eqnarray}
&&G_\Gamma({\bf r}, t,\tsrc) \nonumber \\
&=& \sum_{{\bf x},{\bf x}^{\prime}, {\bf y}^{\prime}}
  \langle 0  |\overline{Q}({\bf x}, t)\Gamma Q({\bf x}+{\bf r}, t) 
  \left(\overline{Q}({\bf x}^{\prime},\tsrc)
  \Gamma Q({\bf y}^{\prime},\tsrc) \right)^{\dagger}
  |0\rangle\nonumber \\
  &=&   \sum_{{\bf x}}\sum_{n} A_n\langle 0|\overline{Q}({\bf x})\Gamma Q({\bf x}+{\bf r}) 
  |n \rangle e^{-M^\Gamma_n(t-\tsrc)},
  \label{eq_correlator} 
\end{eqnarray}
where the gauge field configurations are fixed to the Coulomb gauge 
at both source ($\tsrc$) and sink ($t$) locations.
At source location, both quark and antiquark fields are separately 
averaged in space as wall sources.
The constant amplitude $A_n$ is a matrix element defined as 
$A_n=\sum_{{\bf x}^{\prime}, {\bf y}^{\prime}}
\langle n |\left(\overline{Q}({\bf x}^{\prime})
\Gamma Q({\bf y}^{\prime}) \right)^{\dagger}|0\rangle$.
$M^\Gamma_n$ denotes a rest mass of the $n$-th quarkonium state $|n\rangle$ in
a given $J^{PC}$ channel. Suppose that $|t -\tsrc|/a \gg 1$ is satisfied, the 
four-point correlation function is dominated by the ground state as
\begin{equation}
G_\Gamma({\bf r}, t,\tsrc) 
\xrightarrow{t\gg \tsrc} A_0 \phi_{\Gamma, 0}({\bf r})  
  e^{-M^\Gamma_0(t-\tsrc)},
\end{equation}
where $M_0^{\Gamma}$ is the rest mass of the ground state and the
${\bf r}$-dependent amplitude $\phi_{\Gamma, 0}({\bf r})$
corresponds to the Coulomb gauge BS wave function for the ground state.
For instance, when $\Gamma$ is chosen  
to be $\gamma_5$ for the pseudoscalar (PS) channel $(J^{PC}=0^{-+})$ 
and $\gamma_i$ for the vector (V) channel $(J^{PC}=1^{--})$
in the charm sector, $M_0^{\text{PS}}$ and $M_0^{\text{V}}$ correspond to 
the rest masses of the $\eta_c$ and $J/\psi$ ground states, respectively.
They can be read off from the asymptotic large-time behavior 
of the two-point correlation functions.  Hereafter, we omit the index 0 
from $\phi_{\Gamma, 0}({\bf r})$ and simply call it the BS wave function.

\subsection{Interquark potential defined from BS wave function}
The BS wave function satisfies an effective Schr\"odinger
equation with a nonlocal and energy-independent interquark potential 
$U$~\cite{Ishii:2006ec,Caswell:1978mt,Ikeda:2011bs}:
\begin{equation}
  -\frac{\nabla^2}{2\mu}\phi_\Gamma({\bf r})+
  \int dr'U({\bf r},{\bf r}')\phi_\Gamma({\bf r}')
  =E_\Gamma\phi_\Gamma({\bf r}),
  \label{Eq_schr}
\end{equation}
where the reduced mass $\mu$ of the quarkonium ($\QQbar$) system is given by
a half of the quark kinetic mass  $\mQ$. The energy eigenvalue $E_\Gamma$ of 
the stationary Schr\"odinger equation is supposed to be $M_\Gamma-2\mQ$.
If the relative quark velocity $v=|{\nabla}/\mQ|$ is small as $v \ll 1$, 
the nonlocal potential $U$ can generally expand 
in terms of the velocity $v$ as 
\begin{multline}
 U({\bf r}',{\bf r})=  \{V(r)+V_{\text{S}}(r)\SdotS \\
  +V_{\text{T}}(r)S_{12}+
 V_{\text{LS}}(r){\bf L}\cdot{\bf S} + \mathcal{O}(v^2)\}\delta({\bf r}'-{\bf r}), 
\label{potential}
\end{multline}
 where $S_{12}=(\SQ\cdot\hat{r})(\SQbar\cdot\hat{r})-\SdotS/3$
 with $\hat{r}={\bf r}/r$, ${\bf S}=\SQ+\SQbar$
 and ${\bf L} = {\bf r}\times (-i\nabla)$~\cite{Ishii:2006ec}.
 Here, $V$, $V_{\text{S}}$, $V_{\text{T}}$ and $V_{\text{LS}}$ represent
 the spin-independent central, spin-spin, tensor and spin-orbit potentials, respectively. 
 Remark that the energy dependence on the interquark potential 
 appear at $\Ocal{(v^2)}$.
 For an estimation of the $\Ocal(v^2)$ corrections, 
it is necessary to calculate the BS wave function of higher-lying charmonium states, {\it e.g.}
 the $2S$ charmonium state. Such study is beyond scope of the present paper~\cite{Murano:2010tc}.

The relativistic corrections to the kinetic term have been estimated using 
the relativistic kinematics in Ref.~\cite{Ikeda:2011bs}.
Although the short-range behavior of interquark potential 
is slightly influenced by this modification,
it is indeed small for the heavy quarks such as the charm quark.
Therefore, we here consider the nonrelativistic Schr\"odinger equation with 
spin-dependent corrections up to $\Ocal{(v^2)}$.

In this paper, we focus only on the $S$-wave charmonium states
($\eta_c$ and $J/\psi$).  We perform an appropriate projection 
 with respect to the discrete rotation as 
\begin{equation}
 \phi_{\Gamma}^{A_1^+}({\bf r}) 
     =\frac{1}{24} \sum_{\mathcal{R} \in O_h} \phi_\Gamma (\mathcal{R}^{-1}{\bf r}),
     \label{Eq_phi}
\end{equation}
where $\mathcal{R}$ denotes 24 elements of the cubic point group $O_h$.
This projection provides the BS wave function projected in the $A^{+}_{1}$ representation.
This projected BS wave function corresponds to the $S$-wave 
in continuum theory at low energy~\cite{Luscher:1990ux}. 
We simply denote the $A_1^{+}$ projected
BS wave function by $\phi_{\Gamma}(r)$ hereafter.

The stationary Schr\"odinger equation for the projected 
BS wave function $\phi_{\Gamma}(r)$ is reduced to
 \begin{equation}
  \left\{
    - \frac{\nabla^2}{\mQ}
    +V(r)+\SdotS V_{\text{S}}(r)
  \right\}\phi_{\Gamma}(r)=E_\Gamma \phi_{\Gamma}(r).
  \label{Eq_pot}
\end{equation}
The spin operator $\SdotS$ can be easily 
replaced by expectation values 
$-3/4$ and $1/4$ for the PS and V channels, respectively. 
We here essentially follow usual nonrelativistic potential models,
where the $J/\psi$ state is assumed to be purely composed of the $1S$ wave function.
Within our proposed method, this assumption can be verified 
by evaluating the size of a mixing between $1S$ and $1D$ wave
functions in principle.  

 Both spin-independent and dependent part of 
 the central interquark potentials can be separately evaluated 
 through a linear combination of Eq.~(\ref{Eq_pot}) 
 for PS and V channels as 
 \begin{eqnarray}
   V(r)
   &=& E_{\text{ave}}+\frac{1}{\mQ}\left\{
   \frac{3}{4}\frac{\nabla^2\phi_\text{V}(r)}{\phi_\text{V}(r)}+
    \frac{1}{4}\frac{\nabla^2\phi_\text{PS}(r)}{\phi_\text{PS}(r)}
   \right\} \label{Eq_potC}\\
   V_{\text{S}}(r) 
   &=& E_{\text{hyp}} + \frac{1}{\mQ}\left\{
  \frac{\nabla^2\phi_\text{V}(r)}{\phi_\text{V}(r)} 
  - \frac{\nabla^2\phi_\text{PS}(r)}{\phi_\text{PS}(r)} \right\},\label{Eq_potS}
 \end{eqnarray}
 where $E_{\text{ave}}=M_{\text{ave}}-2\mQ$ 
 and $E_{\text{hyp}}=M_\text{V}-M_\text{PS}$.
 The mass $M_{\text{ave}}$ denotes the spin-averaged mass as  
 $\frac{1}{4}M_\text{PS}+\frac{3}{4}M_\text{V}$.
 The derivative $\nabla^2$ is defined by the discrete Laplacian.
 For other spin-dependent potentials~(the tensor potential 
 $V_{\text{T}}$ and the spin-orbit potential $V_{\text{LS}}$), 
 this approach, in principle, enables us to access them by considering 
 $P$-wave quarkonium states such as the $\chi_{c}$ ($0^{++}$, $1^{++}$) 
 and $h_c$ ($1^{+-}$) states, which must leave contributions 
 of $V_{\text{T}}$ and $V_{\text{LS}}$ to Eq.~(\ref{Eq_pot}).

\subsection{Quark kinetic mass in BS amplitude method}
The definition of the interquark potentials in Eq.~(\ref{Eq_potC}) and (\ref{Eq_potS})
involves {\it unknown information} of the quark mass $m_Q$
that  appears in the kinetic energy term of the effective Schr\"odinger equation~(Eq.~(\ref{Eq_schr}) 
or (\ref{Eq_pot})). 
This is an essential issue on the BS amplitude method when we apply it to the $\QQbar$ system.
Needless to say, the original work, where the BS amplitude method was advocated and applied 
for the nuclear force~\cite{Ishii:2006ec}, 
does not share the same issue since the single-nucleon mass can be 
measured by the standard hadron spectroscopy.

In the initial attempt~\cite{Ikeda:2011bs}, 
the quark kinetic mass $m_Q$ was
approximately evaluated by one-half of the vector quarkonium mass $M_{\rm V}/2$.
However, such an approximate treatment is too crude to define a proper interquark
potential, which could be smoothly connected to the static $\QQbar$ potential
from Wilson loops in the $m_Q \rightarrow \infty$ limit. 
Indeed, it is worth noting that 
the Coulombic binding energy is of order of $m_Q$. 
We may alternatively determine the quark mass from 
the gauge dependent pole mass, which can be measured by 
the quark two-point function in the Landau gauge. 
In this case, we are faced with a difference between the Coulomb and Landau gauges.
In Ref.~\cite{Kawanai:2011xb}, we have solved this issue 
by proposing a novel idea to determine the quark kinetic mass $m_Q$ 
{\it self-consistently within the BS amplitude approach}.  
Let us shortly review the new idea as follows in this subsection.

We start from the spin-spin potential given by Eq.~(\ref{Eq_potS}).
The hyperfine splitting energy, $E_\text{hyp} = M_{\rm V} -M_{\rm PS}$, appeared in Eq.~(\ref{Eq_potS})
can be measured  by the standard hadron spectroscopy.
Under a simple but reasonable assumption:
\begin{equation}
\lim_{r\to\infty} V_{\rm S}(r)=0,
\label{Eq_Assum}
\end{equation}
which implies that there is no long-range correlation and 
no irrelevant constant term in the spin-dependent potential.
Eq.~(\ref{Eq_potS}) is thus rewritten as 
\begin{equation}
 m_Q = \lim_{r\to \infty} \frac{-1}{E_\text{hyp}} \left\{ \frac{\nabla^2\phi_\text{V}(r)}{\phi_\text{V}(r)} 
  - \frac{\nabla^2\phi_\text{PS}(r)}{\phi_\text{PS}(r)} \right\}.
\label{eq_quark_mass}
\end{equation}
This suggests that the quark kinetic mass 
can be read off from the long-distance asymptotic values of 
the difference of quantum kinetic energies (the 2nd derivative of the BS wave function)
in V and PS channels. This idea has been numerically tested, and the 
assumption of Eq.~(\ref{Eq_Assum}) is indeed appropriate in QCD~\cite{Kawanai:2011xb}. 

As a result,  we can self-consistently determine both the spin-independent potential $V(r)$ and 
spin-spin potential $V_{\rm S}(r)$, and also the quark kinetic mass $m_Q$ within a single set of four-point 
correlation functions $G_\Gamma({\bf r},t,\tsrc)$ with $\Gamma=\text{PS}$ and V.

\section{\label{setup}Lattice setup and heavy quarkonium  mass}
\subsection{Quenched gauge ensembles}
\begin{table}
  \caption{
    Simulation paramters of quenched ensembles.
    Lattice spacing $a$ indicates the approximate value with the Sommer scale 
    ($r_0 = 0.5$ fm) input. The table also lists the number of gauge configurations 
    to be analyzed.
    \label{Tab1}
      }
      \begin{ruledtabular}
      \begin{tabular}{ccccccc} 
	Label &$L^3\times T$  &$\beta$ & $a$ [fm]  & $a^{-1}$ [GeV] & $La$ [fm]& Statistics \\ \hline
	FI  &$48^3\times 96$ &6.47 & 0.0469 &4.2 &  2.3  & 100 \\
	ME  &$32^3\times 64$ &6.2  & 0.0677 &2.9 & 2.2  & 150 \\
	CO  &$24^3\times 48$ &6.0  & 0.0931 &2.1 & 2.2  & 300 \\
	LA  &$32^3\times 48$ &6.0  & 0.0931  & 2.1 & 3.0  & 150 \\
      \end{tabular}
      \end{ruledtabular}
\end{table}
In order to understand  the systematics of the BS amplitude method,
we first calculate the interquark potential for the charmonium system
in quenched lattice QCD simulations using several ensembles (three different lattice spacings,
$a\approx 0.093$, $0.068$ and $0.047$ fm, and two different physical volumes, $La \approx 2.2$ and $3.0$ fm). 
The gauge configurations are generated with the single plaquette gauge action.
All lattice spacings are set by the Sommer scale $r_0 = 0.5$~fm~\cite{Sommer:1993ce}.

Three smaller volume ensembles with fixed physical volume ($La \approx 2.2$ fm) 
are mainly employed to test a scaling behavior toward the continuum limit: 
these are the finer lattice ensembles~(FI) on a $24^3 \times 48$ lattice  at $\beta=6.47$,
the medium ones~(ME) on a $32^3 \times 64$ lattice at $\beta=6.2$ 
and  the coarser ones~(CO) on a $24^3 \times 48$ lattice at $\beta=6.0$.
A supplementary data calculated on the larger volume ensembles~(LA) on a $32^3 \times 48$ lattice at $\beta=6.0$
are used for a check of possible finite volume effects.
The number of configurations analyzed is $\Ocal(100$-$300)$.
The gauge configurations are fixed to the Coulomb gauge for calculations of the BS amplitude.
Simulation parameters and the number of sampled gauge configurations
are summarized in Table~\ref{Tab1}.

\subsection{Parameters of RHQ action}
%
%
\begin{table}
  \caption{
  The hopping parameter $\kappa_Q$ and RHQ parameters ($\nu$, $r_s$, $c_B$ and  $c_E$) 
    for the charm quark on all four ensembles.
    \label{Tab2}
      }
      \begin{ruledtabular}
      \begin{tabular}{ccccccc} 
	Label      & $\beta$  & $\kappa_Q$   &$\nu$ & $r_s$ & $c_B$ & $c_E$ \\ \hline
	FI      & 6.47 & 0.11729 & 1.029 & 1.131 & 1.700 & 1.562 \\
	ME    & 6.2  & 0.11035 & 1.050 & 1.185 & 1.898 & 1.710 \\
	CO, LA  & 6.0  & 0.10072 & 1.088 & 1.273 & 2.194 & 1.932 \\
      \end{tabular}
      \end{ruledtabular}
\end{table}
The heavy quark propagators are computed using the RHQ
action that has five parameters $\kappa_Q$, $\nu$, $r_s$, $c_B$ and  $c_E$~\cite{Aoki:2001ra}.
The RHQ action used here is a variant of the Fermilab approach, and 
can control large discretization errors introduced by large quark mass~\cite{ElKhadra:1996mp}
(See also~Refs.~\cite{Christ:2006us,Lin:2006ur,Aoki:2012xaa}).
The hopping parameter is chosen to reproduce the experimental 
spin-averaged mass of $1S$ charmonium states $M_\text{ave}^\text{exp}=3.0678(3)$ GeV~\cite{Beringer:1900zz}
at each lattice spacing.
The five RHQ parameters are basically determined by one-loop perturbative 
calculations~\cite{Kayaba:2006cg}.
The parameter sets of the RHQ action in quenched simulations at three lattice spacings
are summarized in Table~\ref{Tab2}.

We calculate quark propagators with a wall source.
Dirichlet boundary conditions are imposed for the time direction.
In order to investigate the energy-momentum dispersion relation,
we also employ a gauge invariant Gaussian-smeared source~\cite{Gusken:1989qx}
for the standard two-point correlation function with four finite momenta:
$a {\bf p}  = 2\pi/L \times (1,0,0)$, $(1,1,0)$, $(1,1,1)$ and $(2,0,0)$.

\subsection{\label{sec:effective_mass}Effective mass from two point function}
\begin{table*}
  \caption{Fitted masses of $1S$ charmonium states, 
    their spin-averaged masses and hyperfine splitting
    energies obtained with the same fit range on all four ensembles.    
    Results are tabulated in units of GeV. 
    The hopping parameter $\kappa$ on each ensembles is chosen
    approximately to reproduce the experimental spin-averaged
    mass of 1S charmonium states $M_\text{ave}^{\rm exp}=3.0678(3)$~GeV~\cite{Beringer:1900zz}.
    \label{Tab3}
     }
      \begin{ruledtabular}
	\small
      \begin{tabular}{cclclcll} 
	Label &  fit range 
	& \multicolumn{2}{c}{$\eta_c$ mass} & \multicolumn{2}{c}{$J/\psi$ mass}
	& spin-averaged  mass 	& hyperfine splitting energy  \\
	& $[t_\text{min}/a:t_\text{max}/a]$ 
	& $M_{\eta_c}$ [GeV] & $\chi^2/{\rm d.o.f.}$  & $M_{J/\psi}$  [GeV]& $\chi^2/{\rm d.o.f.}$ 
	& $M_{\rm ave}$ [GeV]& $E_{\rm hyp}$ [GeV] \\[2pt] \hline 
        FI  & $[54:72]$ & 3.0121(14) & 0.66 &  3.0861(22) & 0.62 & 3.0676(20) & 0.0741(11)  \\
        ME  & $[37:50]$ & 3.0188(10) & 0.55 &  3.0980(18) & 0.93 & 3.0783(15) & 0.0773(11) \\
        CO  & $[27:36]$ & 3.0126(8)  & 1.65 &  3.0923(13) & 1.02 & 3.0724(11) & 0.0795(8)\\
        LA  & $[30:39]$ & 3.0120(8)  & 0.98 &  3.0907(13) & 0.75 & 3.0710(10) & 0.0790(8) \\
      \end{tabular}
      \end{ruledtabular}

\end{table*}
 \begin{figure}
   \centering
   \includegraphics[width=.49\textwidth]{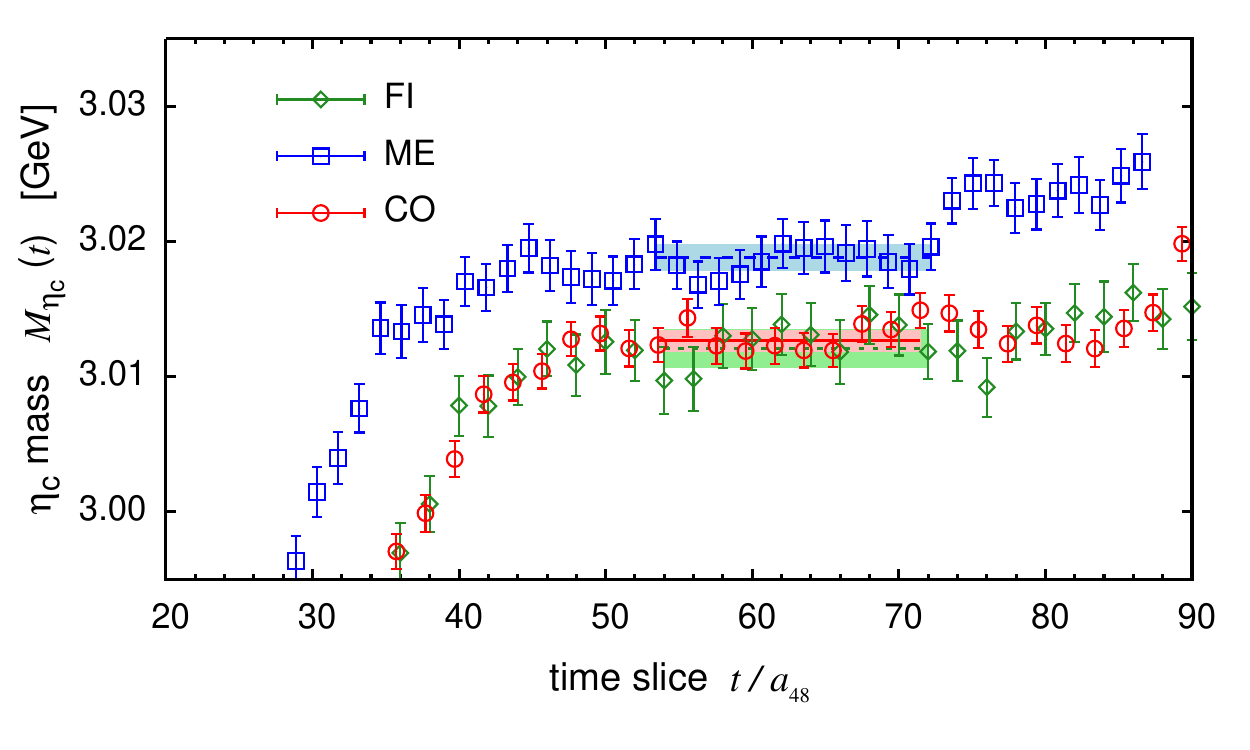}
   \includegraphics[width=.49\textwidth]{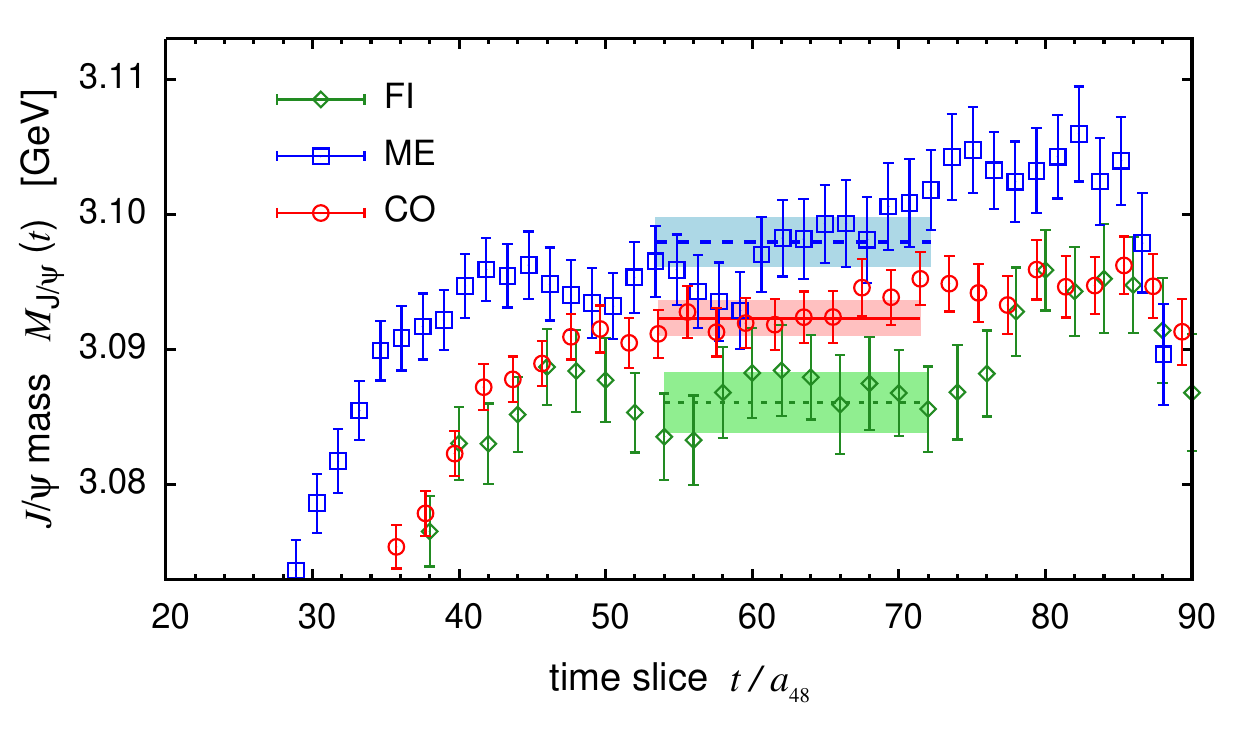}
   \caption{Effective mass plots for the $\eta_c$~(upper) and $J/\psi$~(lower).
    Three different symbols indicate results obtained
    on the CO (circles), ME (squares) and FI (diamonds) ensembles.
    The horizontal axis is plotted in units of $a_{48}$, which is the lattice spacing of
    the FI ensembles. 
    Solid, dashed and dotted lines represent the fit results for effective mass 
    calculated on the CO, ME and FI ensembles, respectively.
    Shaded bands denote fit ranges and statistical errors estimated by the jackknife method.
  }
  \label{effmass}
 \end{figure}

The mass $M_\Gamma$ of the charmonium states ($\Gamma=$ PS and V) is extracted 
by the two-point function.
When a separation between a quark and an anti-quark at the sink is set to be zero (${\bf r}=0$), 
the four-point correlation functions  $G_\Gamma({\bf r},t,\tsrc)$ defined in Eq.~(\ref{eq_correlator}) 
simply reduces the usual two-point function  $G_\Gamma(t,\tsrc)$
with a wall source. The effective mass functions are then defined as 
\begin{equation}
 M_\Gamma(t) = \log\frac{G_\Gamma(t,\tsrc)}{G_\Gamma(t+1,\tsrc)}.
\end{equation}

Fig.~\ref{effmass} shows the effective mass plots of 
the $1S$ charmonium states ($\eta_c$ and  $J/\psi$), 
calculated on three ensembles~({FI, ME, CO}).
Each effective mass plot shows a reasonable plateau.  
We estimate the $\eta_c$ and $J/\psi$ masses by a constant fit to the plateaus over time ranges shown in Table~\ref{effmass}.
A correlation between masses measured at various time slices is took  into account
by using a covariance matrix in the constant fit.
A inversion of covariance matrix is performed once for average and used for each jackknife block.
The statistical uncertainties indicated by shaded bands in Fig.~\ref{effmass}
is estimated by the jackknife method.
For all ensembles, we basically take similar fitting ranges in the same units, 
indicated by shaded bands in Fig.~\ref{effmass}.
All fit results are summarized in Table~\ref{Tab3}.
Also, values of spin-averaged mass $M_\text{ave}$ and 
hyperfine splitting energy $E_\text{hyp}$ are quoted in Table~\ref{Tab3}.
Note that we simply neglect the disconnected diagrams in 
calculations of both four-point and two-point correlation
 functions for the $\eta_c$ and $J/\psi$ states in our simulations.

We observe that on the FI ensembles the data of four-point and two-point 
correlation functions at different time slices are highly correlated. 
This strong correlation between the time slices becomes more pronounced
when we calculate the interquark potential from the BS wave function.
In the analysis of the interquark potential, we have to somehow reduce the correlation, 
which makes the covariance matrix singular, in order to get a reasonable value of 
$\chi^2$/d.o.f. during the fitting. Therefore we will use the data points at even number time slices 
for evaluation of the BS wave function on the FI ensembles. Note that the effective mass 
for the FI ensembles was evaluated only with even number time slices to perform a consistent analysis
(see Fig.~\ref{effmass}, where the FI data points are appeared only at even number time slices).

The spin-averaged masses measured on the ME and CO (LA) ensembles slightly deviate from 
experimental data~(see Table \ref{Tab3}). This implies that our
calibration for the hopping parameters of the charm sector is not precise enough.
Then, strictly speaking, a systematic uncertainty due to tuning the charm quark mass 
is larger than the statistical one. However its accuracy is still enough to study the interquark
potential for the charmonium in this quenched studies. 
As we will discuss later, although the discrepancy among the spin-averaged masses given 
at three lattice spacings is kept less than one percent, 
the resulting quark kinetic masses are fairly consistent with each other albeit with
rather large statistical errors.

 \begin{figure}
   \centering
   \includegraphics[width=.49\textwidth]{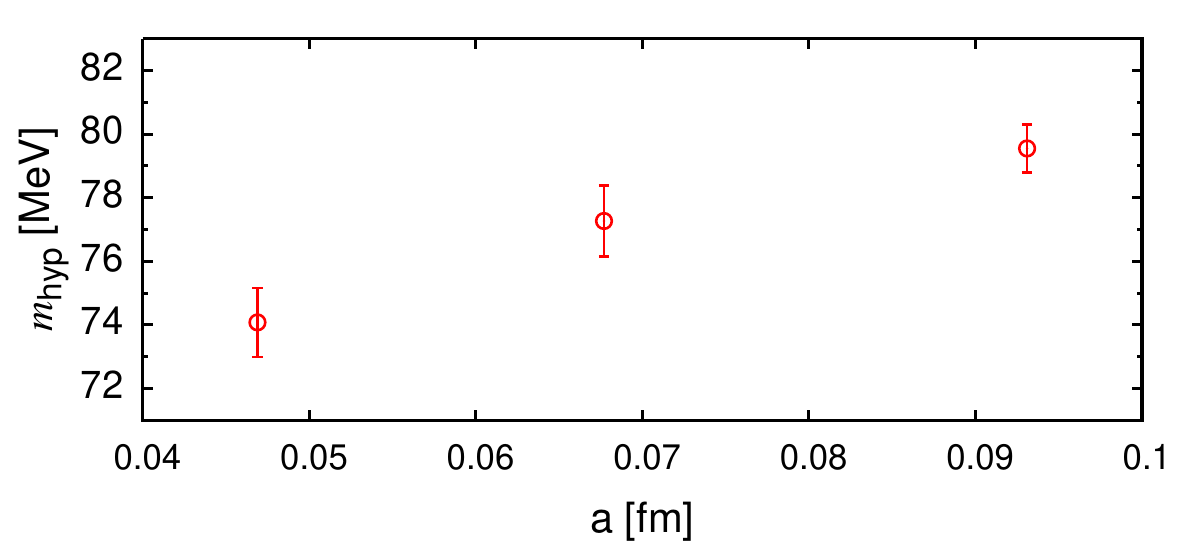}
   \caption{The lattice spacing dependence of hyperfine splitting energies
    calculated on the FI, ME and CO ensembles.
    Results are shown in units of MeV as a function of lattice spacing in units of fm.
  }
  \label{hyperfine_E}
 \end{figure}
For the hyperfine splitting energy, 
results obtained from our quenched simulations 
reproduce only $65-70$~\% of experimental value 
$M_{\rm hyp}^{\rm exp} = 113.2(7)~{\rm MeV}$~(See also Ref.~\cite{Kawanai:2011jt}).
As shown in Fig.~\ref{hyperfine_E}, the data points exhibit a slight linear dependence of the lattice spacing. 
We consider that the lattice spacing dependence of the hyperfine splitting energy
is mainly caused by a remnant ${\cal O}(a)$ discretization error,
rather than the issues related to calibration of the precise measurement at the charm quark mass,
since our RHQ action with one-loop perturbative coefficients do not fully improve
the leading discretization error. 
From this observation, we speculate that the discretization effect would be non-negligible 
for the spin-spin potential, which is highly sensitive to the hyperfine splitting energy by the definition 
given in Eq.~(\ref{Eq_potS}).

 \begin{figure*}
   \centering
   \includegraphics[width=.32\textwidth]{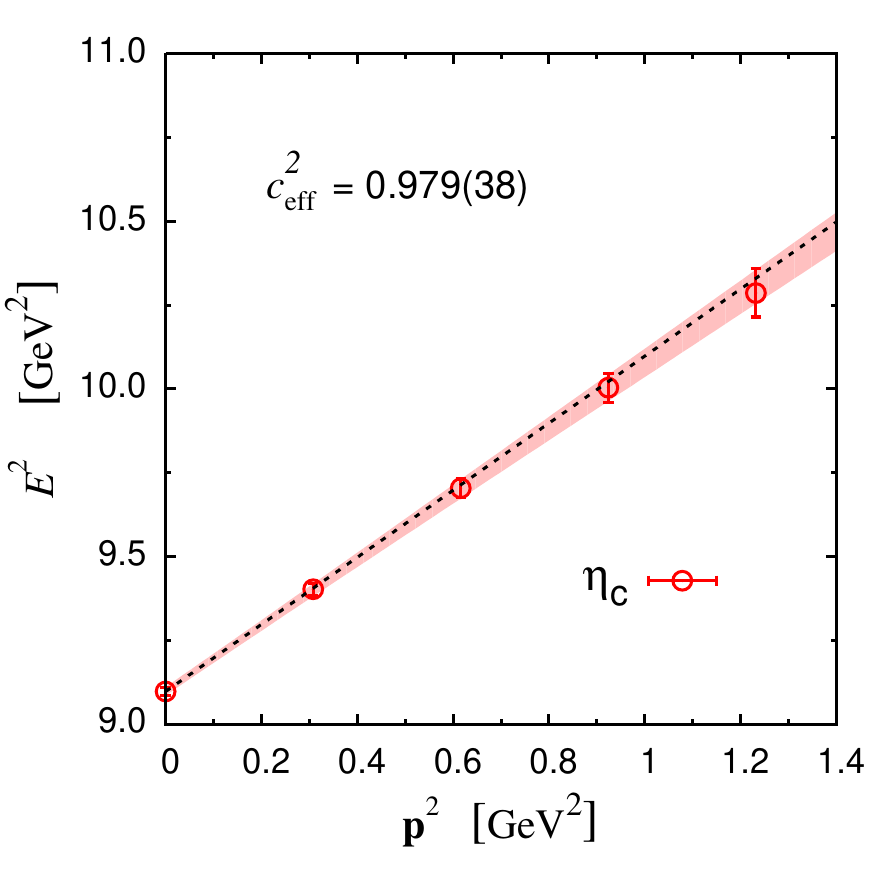}
   \includegraphics[width=.32\textwidth]{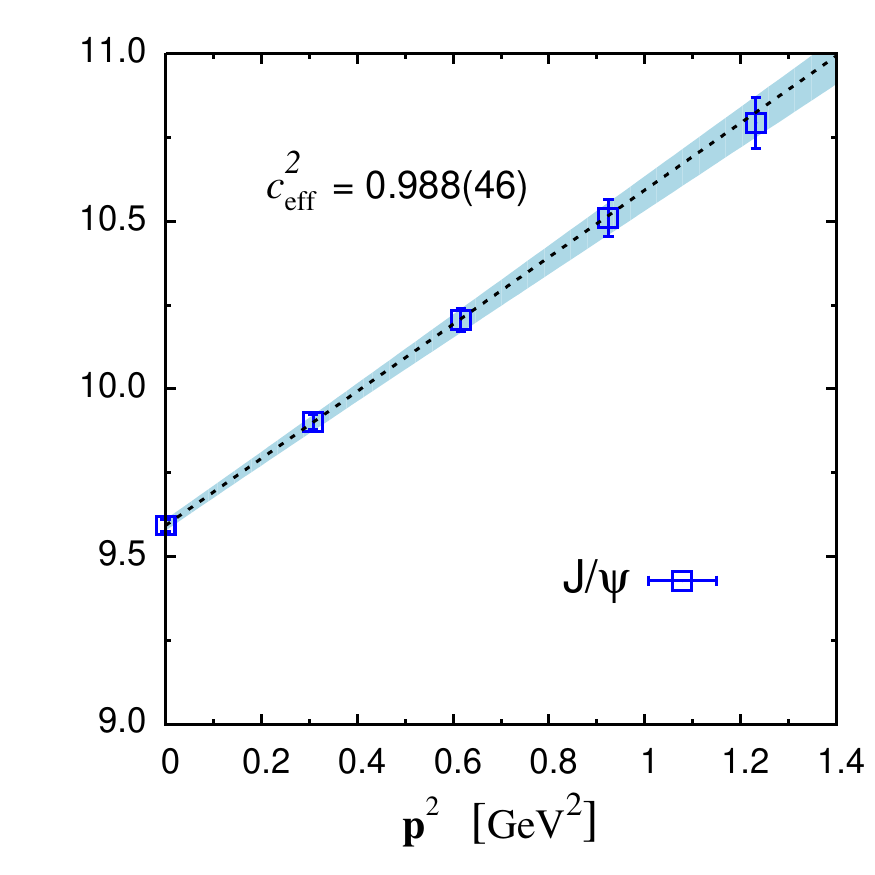}
   \includegraphics[width=.32\textwidth]{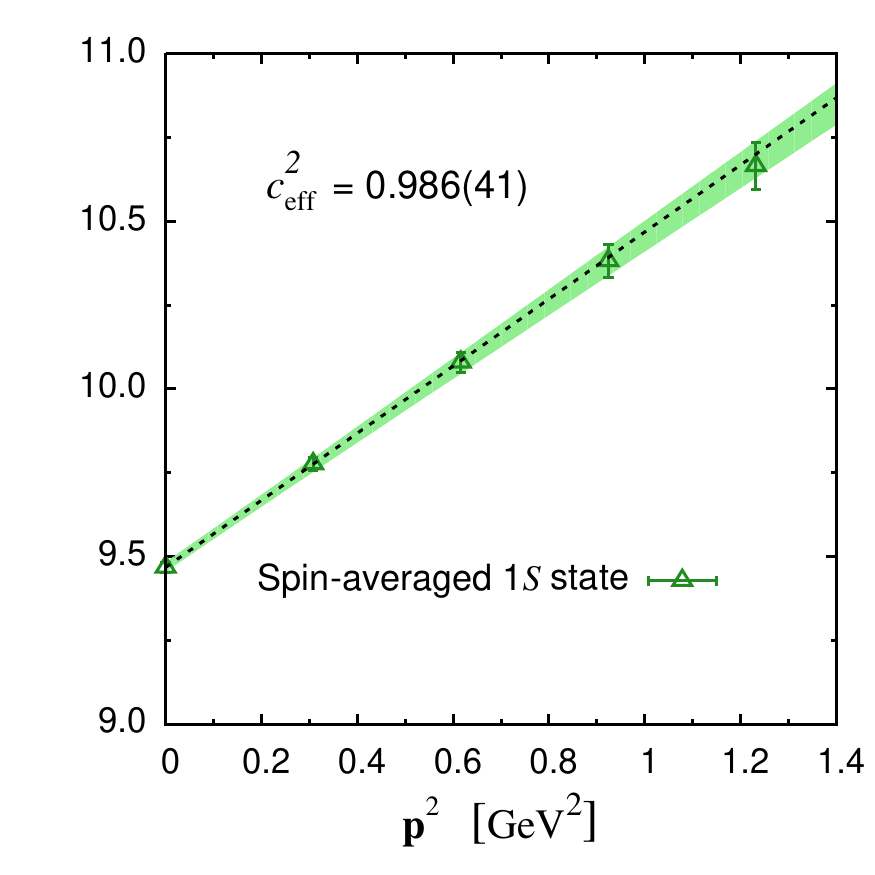}
   \caption{Check of the energy-momentum dispersion relation
    $E^2({{\bf p}}^2)= M^2 + c_\text{eff}^2{{\bf p}}^2$
    for the $\eta_c$~(left), $J/\psi$~(center) and spin-averaged $1S$ state~(right)
    calculated on the CO ensembles, as typical examples.
    By	 the linear fit to data points calculated with various spatial momenta
    including zero momentum, the effective speed of light is obtained.
    Values of the squared effective speed of light $c_{\rm eff}^2$
    are quoted in each panel. Shaded bands indicate statistical uncertainties
    in fitting, estimated by the jackknife method.
    For comparison, the continuum dispersion relation ($c_\text{eff}^2 = 1$) 
    is denoted as the dashed lines in each panel.
}
  \label{dispersion}
 \end{figure*}
In the continuum theory, a relativistic particle, of which the rest mass is $m$, 
moving with spatial momentum ${\bf p}$ obeys 
the energy-momentum dispersion relation as
 $E^2 = m^2 + {\bf p}^2$. However, in lattice QCD, the dispersion relation deviates
 from the continuum one due to the presence of lattice discretization corrections  as
 \begin{eqnarray}
(aE)^2 = (am)^2 + c_{\rm eff}^2(a{\bf p})^2+c'|a{\bf p}|^4 + \mathcal{O}(a^6), 
\end{eqnarray}
where the spatial momentum is given by 
$a{\bf p}=\frac{2\pi\mathbf{n}}{L},  \  \mathbf{n} \in \mathbb{Z}^3$
in a finite $L^3$ box with periodic boundary conditions.
A coefficient $c_{\rm eff}^2$ appearing in the second term is squared effective speed of light.
In the continuum limit $a\to 0$,  $c_{\rm eff}$ should be unity and
higher order corrections vanish. 
If the discretization effect due to finite lattice spacing is well under control 
by using an improved action, $c_{\rm eff}^2$ is supposed to remain approximately unity.

Fig.~\ref{dispersion} shows the energy-momentum dispersion relations
for the $\eta_c$ and $J/\psi$ states, and their spin-averaged one 
on the CO ensembles as a typical example.
Our data up to spatial momenta of ${\bf n}^2 = 4$
well reproduces the continuum dispersion relation and resulting $c_{\rm eff}^2$ 
is consistent with unity within error bars. The other ensembles also provide the similar results.

\section{\label{main} Determination of interquark potential}
\subsection{\label{sec:wavefunc}$\QQbar$ BS wave function}
 \begin{figure}
   \centering
   \includegraphics[width=.49\textwidth]{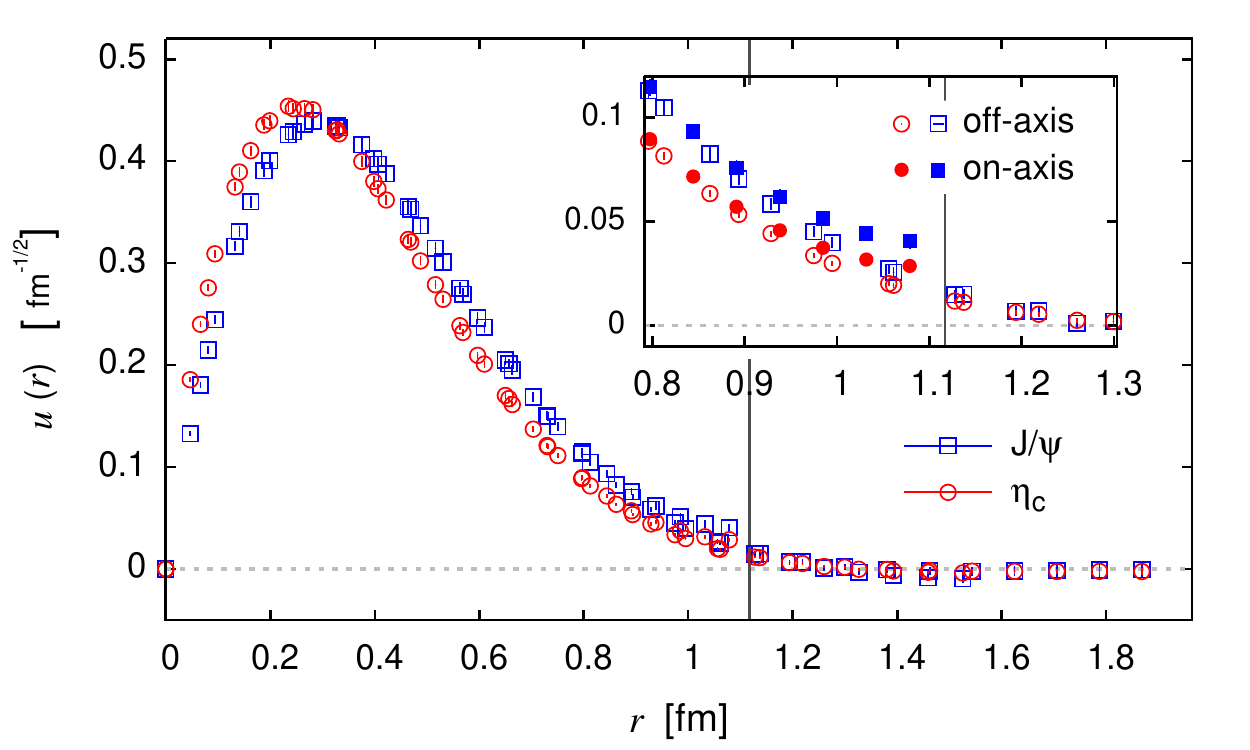}
   \caption{
    The $\QQbar$ BS wave functions of the $\eta_c$~(circles)
    and $J/\psi$~(squares) states calculated
    using the FI ensembles~($a\approx 0.047$~fm), shown as a function of the spatial distance~$r$.
    The data points are taken at $\bf r$ vectors, which are multiples of three directions,~$(1,0,0)$, $(1,1,0)$ and $(1,1,1)$.
    A vertical solid line marks the position of a half of the lattice size~($La/2\approx 1.1$~fm).
    The inset shows a magnified view of the wave functions around $r\approx La/2$
    and filled symbols in the inset represent the data points taken along the on-axis direction.
    }
  \label{wavefunction1}
 \end{figure}
 \begin{figure}
   \centering
   \includegraphics[width=.49\textwidth]{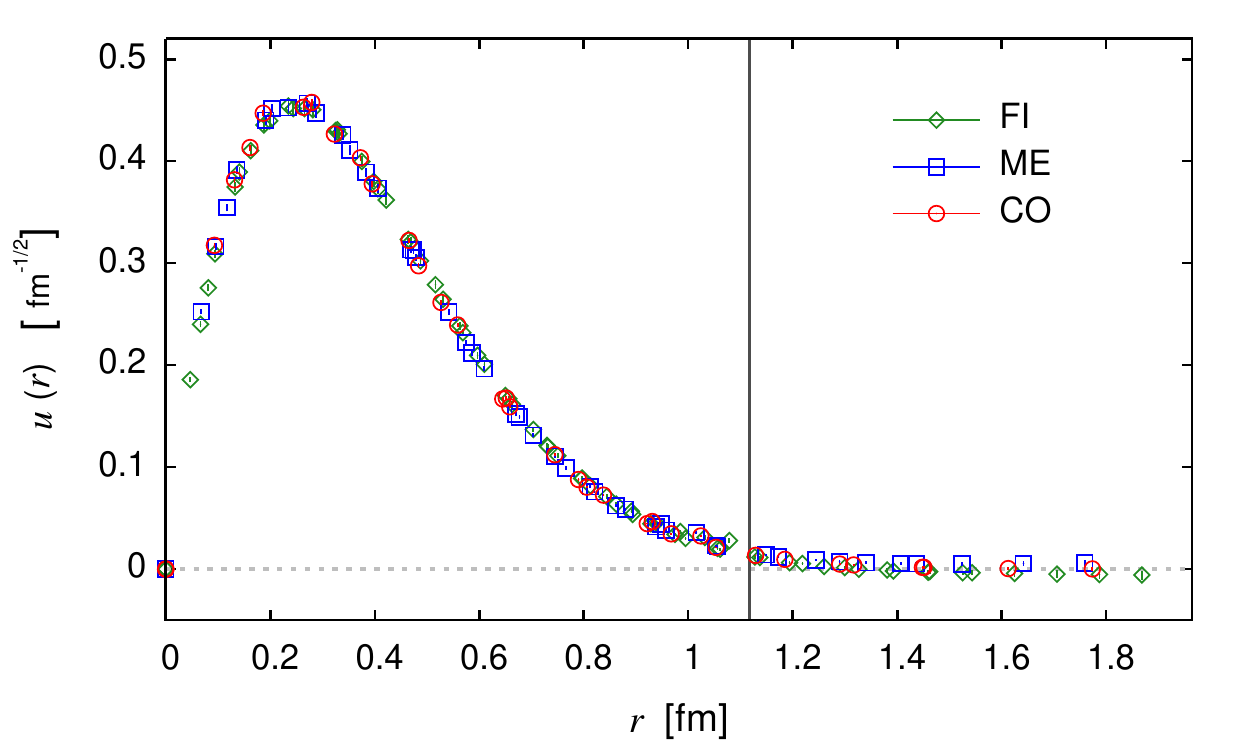}
  \caption{
    The $\QQbar$ BS wave functions of the $\eta_c$ state calculated on 
    the FI, ME and CO ensembles. 
    A vertical solid line marks the position of a half of the lattice size~($La/2\approx 1.1$~fm).          
  }
  \label{wavefunction2}
 \end{figure}
In Fig.~\ref{wavefunction1}, 
we show the reduced $\QQbar$ BS wave functions 
$u_\Gamma(r) =r\tilde{\phi}_\Gamma({\bf r})$ of $1S$ charmonium 
states ($\eta_c$ and $J/\psi$ states), calculated on the FI ensembles. 
The normalized $\QQbar$ BS wave function with the definition given in Eq.~(\ref{eq_phi_gauge_inv}) can be evaluated by
the following ratio of four-point correlation functions $G_\Gamma({\bf r},t, \tsrc)$ 
at large Euclidean time:
\begin{equation}
  \tilde{\phi}_\Gamma({\bf r}) =
  \frac{\phi_\Gamma({\bf r})}{\displaystyle \sum_{\bf r} \{\phi_\Gamma({\bf r})\}^2} 
   = \lim_{|t-\tsrc|\rightarrow \infty}   \frac{G_\Gamma({\bf r},t, \tsrc)}{\displaystyle \sum_{\bf r} \{G_\Gamma({\bf r},t, \tsrc)\}^2}.
\label{eq:BS_wavefunction} 
\end{equation}
We take an average of this ratio with respect to the time slice by a weighted sum 
in the range, where the effective mass of the $1S$ charmonium states exhibits a clear 
plateau behavior.
Here, the normalized wave function $\tilde{\phi}_\Gamma({r})$ satisfies a 
condition $\sum \tilde{\phi}_{\Gamma}^2 = 1$.
We use the reduced wave function $u_\Gamma(r)$ for displaying the spatial distribution of 
the BS wave function. We focus on data points taken at $\bf r$ vectors, which are multiples 
of three directions, $(1,0,0)$~(on-axis), $(1,1,0)$~(off-axis I) 
and $(1,1,1)$~(off-axis II).

As shown in Fig.~\ref{wavefunction1}, the $\QQbar$ BS wave function projected 
in the $A_1^+$ representation, which corresponds to the $S$-wave in the continuum 
theory, is certainly isotropic as was expected. In general, the breaking of rotational symmetry
is one of major artifacts associated with the discretization error. 
However, there is no sufficient difference between 
the $\QQbar$ BS wave functions calculated along three different directions. 
It suggests that the discretization effect due to finite lattice spacing would 
be considerably small. Indeed, the $\QQbar$ BS wave function of the $\eta_c$ state 
shows a good scaling behavior as shown in Fig.~\ref{wavefunction2}.
All data of the $\eta_c$ wave function obtained from three ensembles~({LI, ME, CO}) 
clearly fall onto a single curve. Nothing changes for the $J/\psi$ wave function.

Fig.~\ref{wavefunction1} and \ref{wavefunction2} show that 
the $\QQbar$ BS wave functions of $1S$ charmonium states vanish for $r \alt$ 1~fm and
eventually fit into the lattice volume utilized here.
Such localized wave functions indicate that the $1S$ charmonium states are  bound states. 
Therefore, the finite volume effect on the interquark potential is expected to be small,
 and the spatial extent of the present lattice size
($La\approx 2.2$~fm) is likely to be large enough to study the $1S$ charmonium states. 
However, there is still some caveat for the on-axis data.
The vanishing point $r\sim 1$~fm is very close to a half of the spatial extent of the present lattice
size, which is depicted as a solid vertical line in Fig.~\ref{wavefunction1} and \ref{wavefunction2}.
A wrap-round effect would be set in the on-axis direction near the spatial boundary.
In fact, the on-axis data marginally deviates from the off-axis data
at around $r\sim 1$~fm~(See the inset of Fig.~\ref{wavefunction1}).

\subsection{\label{sec:laplacian_op}Discrete Laplacian operator}
\begin{figure}
   \centering
   \includegraphics[width=.49\textwidth]{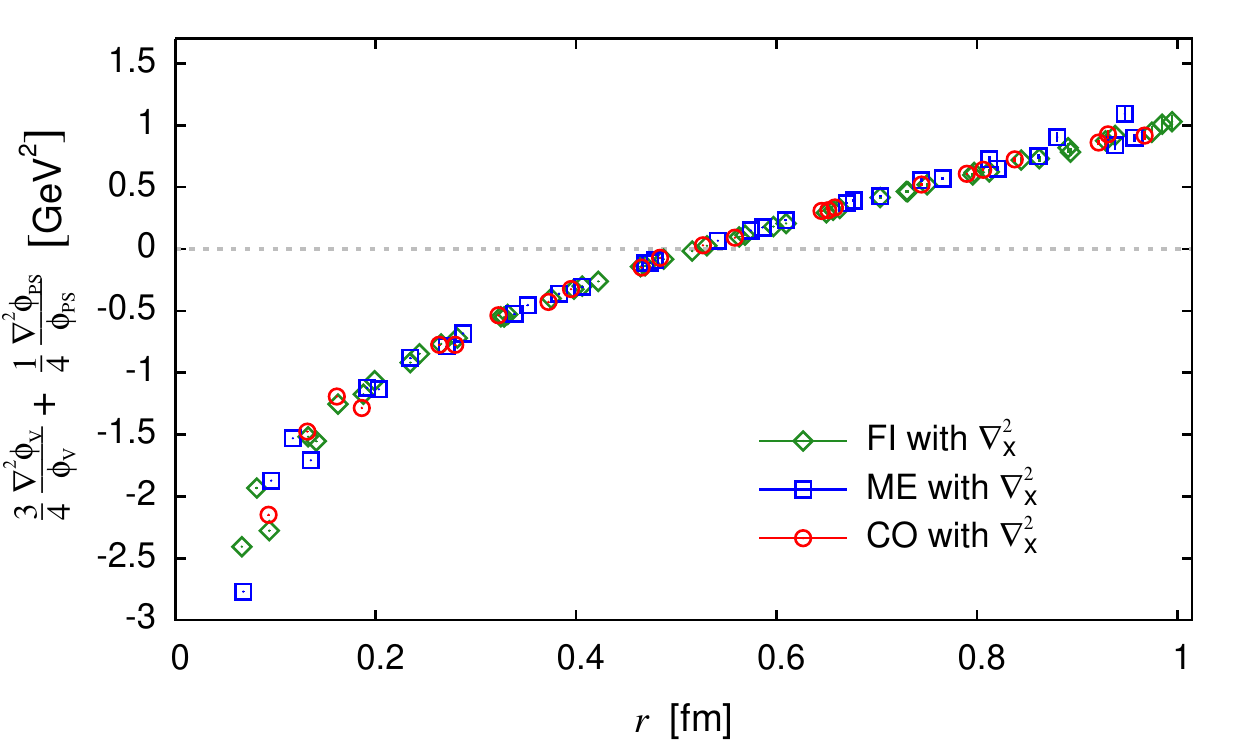}
   \includegraphics[width=.49\textwidth]{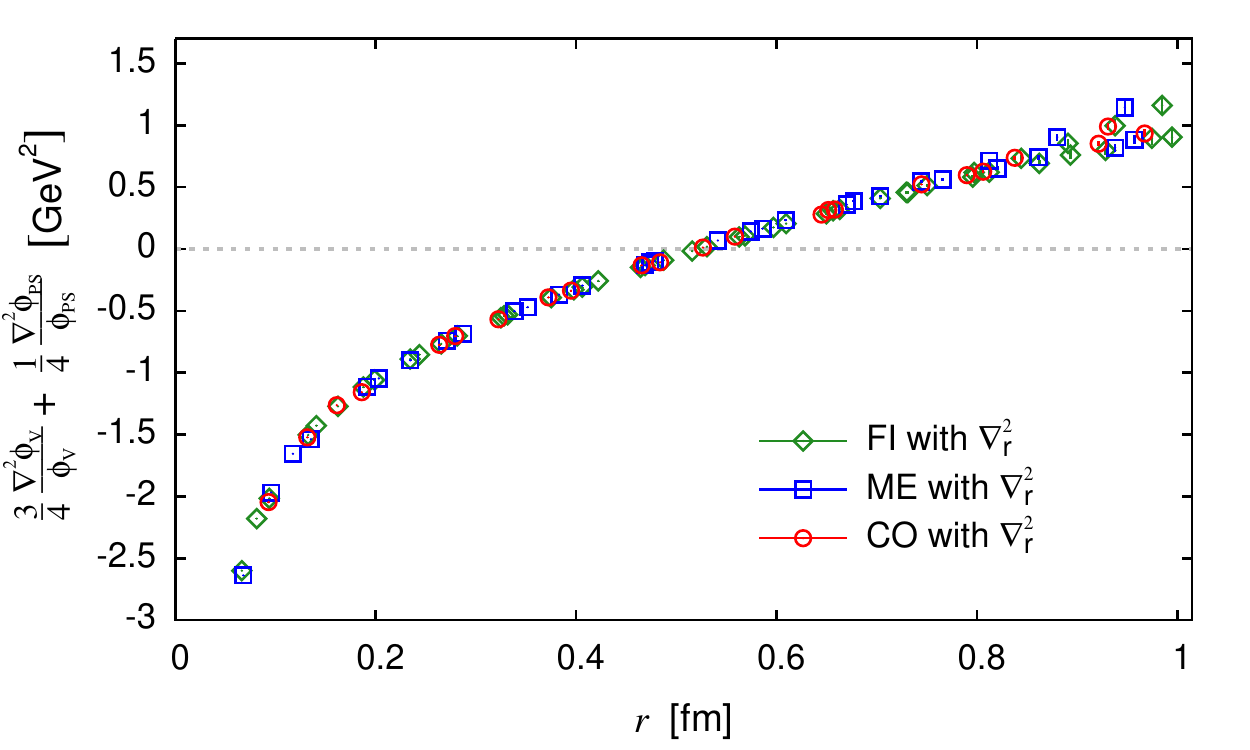}
  \caption{
    The spin-averaged ratios 
    $\frac{3}{4}\nabla_{\rm lat}^2\phi_{\rm V}/\phi_{\rm V}+\frac{1}{4}\nabla_{\rm lat}^2\phi_{\rm PS}/\phi_{\rm PS}$
    as functions of the spatial distance $r$, which are calculated
    with the discrete $x$-Laplacian $\nabla_{\bf x}^2$~(upper) and discrete $r$-Laplacian $\nabla_{\bf r}^2$~(lower)
    operators.
    Three different symbols denote results obtained from three different ensembles:
    the CO (circles), ME (squares) and FI (diamonds) ensembles.
    \label{laplacian}
  }
 \end{figure}
We next discuss choices of the discrete Laplacian operator $\nabla^2_{\rm lat}$, 
which is  built in the definition of the interquark potential.
The discrete Laplacian operator on lattice can be naively defined
with nearest neighbor points in the Cartesian coordinate system as below, called {\it x-Laplacian} in this paper:
\begin{eqnarray}
   \nabla_{\bf x}^2 \phi_\Gamma({\bf r})
  &=& \sum_{\hat{\bf i}=\hat{x},\hat{y},\hat{z}} \frac{1}{a^2}
  \{\phi_\Gamma({\bf r} + \hat{\bf i})a + \phi_\Gamma({\bf r}+\hat{\bf i}a) -2\phi_\Gamma({\bf r})\} \nonumber \\
  &=& \nabla_{\rm cont}^2 \phi_\Gamma({\bf r}) +\Ocal(a^2),
\label{eq_laplacian_cartecian}
\end{eqnarray}
where $\nabla_{\rm cont}^2$ is the continuum Laplacian operator.
A discretization error introduced by the discrete derivative operator starts at $\Ocal(a^2)$.

In order to clarify the systematic uncertainties of the discrete Laplacian,
we focus on a spin-averaged ratio
\begin{equation}
  \Vcal(r)= \frac{3}{4}\nabla^2_{\rm lat}\phi_{\rm V}/\phi_{\rm V}+\frac{1}{4}\nabla^2_{\rm lat}\phi_{\rm PS}/\phi_{\rm PS} , 
\end{equation}
which is associated with the spin-independent interquark
potential apart from the vertical scale and offset.
This spin-averaged ratio is suitable for understanding the systematic uncertainty 
on the discrete Laplacian.
Statistical and systematic uncertainties of $\Vcal(r)$ are
relatively small due to absence of the quark kinetic mass, whose determination
introduces large statistical fluctuation, in comparison to the potential itself.
In other wards, this spin-averaged ratio is independent of the definition of the quark mass.

The upper panel of Fig.~\ref{laplacian} shows the spin-averaged ratios $\Vcal(r)$ 
calculated with the $x$-Laplacian using the {FI}, {ME} and {CO} ensembles.
Although the ratios in the upper panel of Fig.~\ref{laplacian} show more or less 
the same scaling behavior as found in the $\QQbar$ BS wave function, 
some multiple-valuedness, which represents the rotational symmetry breaking, 
appears at short distances and also at long distances. 
Near the spatial boundary $r\agt 0.9$~fm, this unexpected sign of the rotational symmetry 
breaking could be explained by the finite volume effect. 
In practice, we naturally have a difficulty to obtain reliable data at long distances
because the $\QQbar$ wave functions of $1S$ charmonium state
are quickly dumped~(Fig.~\ref{wavefunction1} and \ref{wavefunction2}) 
and signal-to-noise ratio turns out to be worse 
in the ratio $\nabla^2\phi_\Gamma/\phi_\Gamma$.
However, we have already seen in Fig.~\ref{wavefunction1} that the on-axis data slightly 
deviates from the off-axis data near the spatial boundary in the BS wave function. 
Such small finite size effect should be inherited in the interquark potential. 

On the other hand, the multivalued spin-averaged ratio appeared in $r\alt 0.3$ fm is 
inconsistent with no sign of the rotational symmetry breaking in the BS wave function.
This implies that the multiple-valuedness appeared near the origin in the spin-averaged ratios
is mainly stemming from the discretization artifact of the Laplacian operator.

To reduce the possible discretization error at short distances,
we try to consider the following discrete Laplacian operator defined 
in the discrete polar coordinate,
called {\it r-Laplacian}:
\begin{eqnarray}
\nabla_{\bf r}^2 \phi_\Gamma(r)
&=& \frac{2}{r}\frac{\phi_\Gamma(r+\tilde{a})-\phi_\Gamma(r-\tilde{a})}{2\tilde{a}} \nonumber \\
& & + \frac{\phi_\Gamma(r+\tilde{a})+\phi_\Gamma(r-\tilde{a})-2\phi_\Gamma}{\tilde{a}^2} \nonumber \nonumber \\
&=& \frac{2}{r}\frac{\partial}{\partial r}\phi_\Gamma(r)+
\frac{\partial^2}{\partial r^2}\phi_\Gamma(r)
+\Ocal(\tilde{a}^2),
\label{eq_laplacian_polar}
\end{eqnarray}
where $r$ is the absolute value of the relative distance as $r = |{\bf r}|=\sqrt{x^2+y^2+z^2}$ and
$\tilde{a}$ is a distance between grid points along differentiate directions.
We compute the ratio $\Vcal(r)$ with the polar Laplacian in
three directions: the on-axis, off-axis~{I} and off-axis~{II}, where
the effective grid spacings correspond to $\tilde{a} = a, \sqrt{2}a, \sqrt{3}a$, respectively.
Note here that the discretization errors induced by $\nabla_{\bf r}^2$ in the off-axis~I and II directions
are two and three time as much as in the on-axis direction, respectively.

In Eq.~(\ref{eq_laplacian_polar}), we assume that the $\QQbar$ BS wave 
function $\phi_\Gamma({\bf r})$ depends only on the distance $r$, 
namely $\phi_\Gamma({\bf r})$ is isotropic.
This is a reasonable assumption for the data shown in Fig.~\ref{wavefunction2}.
Small, but visible effects of rotational symmetry breaking on  $\phi_\Gamma({\bf r})$  is simply encoded 
  into discretization effects on the ratio $\Vcal(r)$.
  They must vanishes at the continuum limits $a\to 0$ and infinite volume limits $L \to \infty$.

The derivative term of $\nabla_{\rm lat}^2\phi_\Gamma(r)$ 
evaluated with both discrete Laplacian operators, $\nabla_{\bf x}^2$ and $\nabla_{\bf r}^2$,  
must essentially give the same answer  in $a\to 0$ and  $L \to \infty$.

The spin-averaged ratio $\Vcal(r)$ calculated with $\nabla_{\bf r}$ is shown in the lower panel of Fig.~\ref{laplacian}.
A shape of the ratio obtained with the polar Laplacian is 
highly improved to satisfy the single-valuedness at short distances. 
Similar to the $\QQbar$ BS wave function,
the data points of the spin-averaged ratio calculated on three different ensembles
fall onto a single curve at short distances.
The rotational symmetry is also effectively recovered.

These results suggest us that the discrete polar Laplacian operator is 
better than the naive one to evaluate the interquark potential 
from the $S$-wave $\QQbar$ BS wave function.
On the other hand, the rotational symmetry breaking observed
at long distances due to the finite volume is not cured, or rather slightly enhanced.
In this work, the $r$-Laplacian operator $\nabla^2_{\bf r}$
is used for the 2nd derivatives $\nabla^2$.
Then, the subscript ${\bf r}$ on $\nabla^2_{\bf r}$ is omitted hereafter.

\subsection{\label{sec:time_average} Time average}
\begin{figure*}
   \centering
   \includegraphics[width=.49\textwidth]{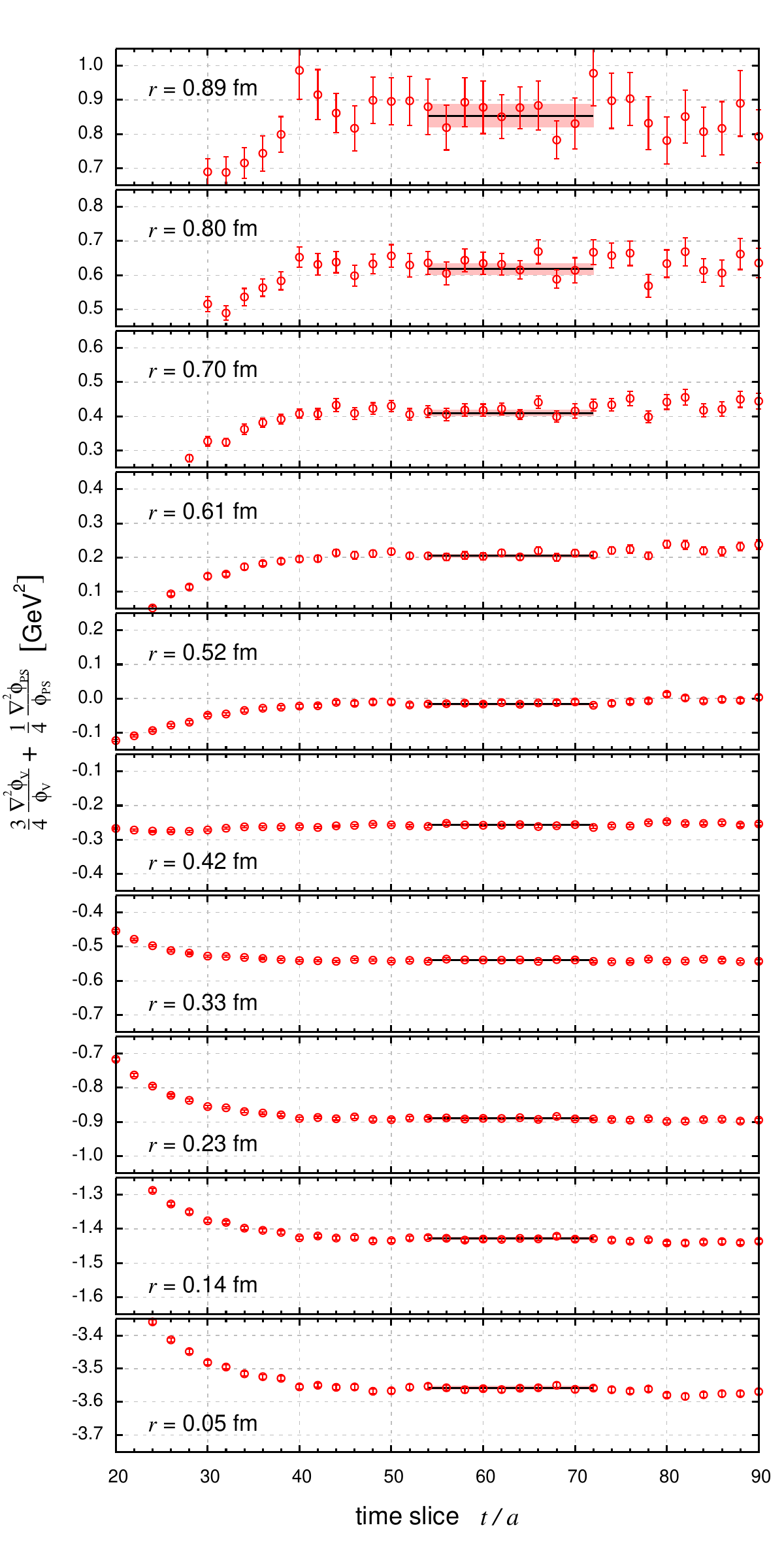}
   \includegraphics[width=.49\textwidth]{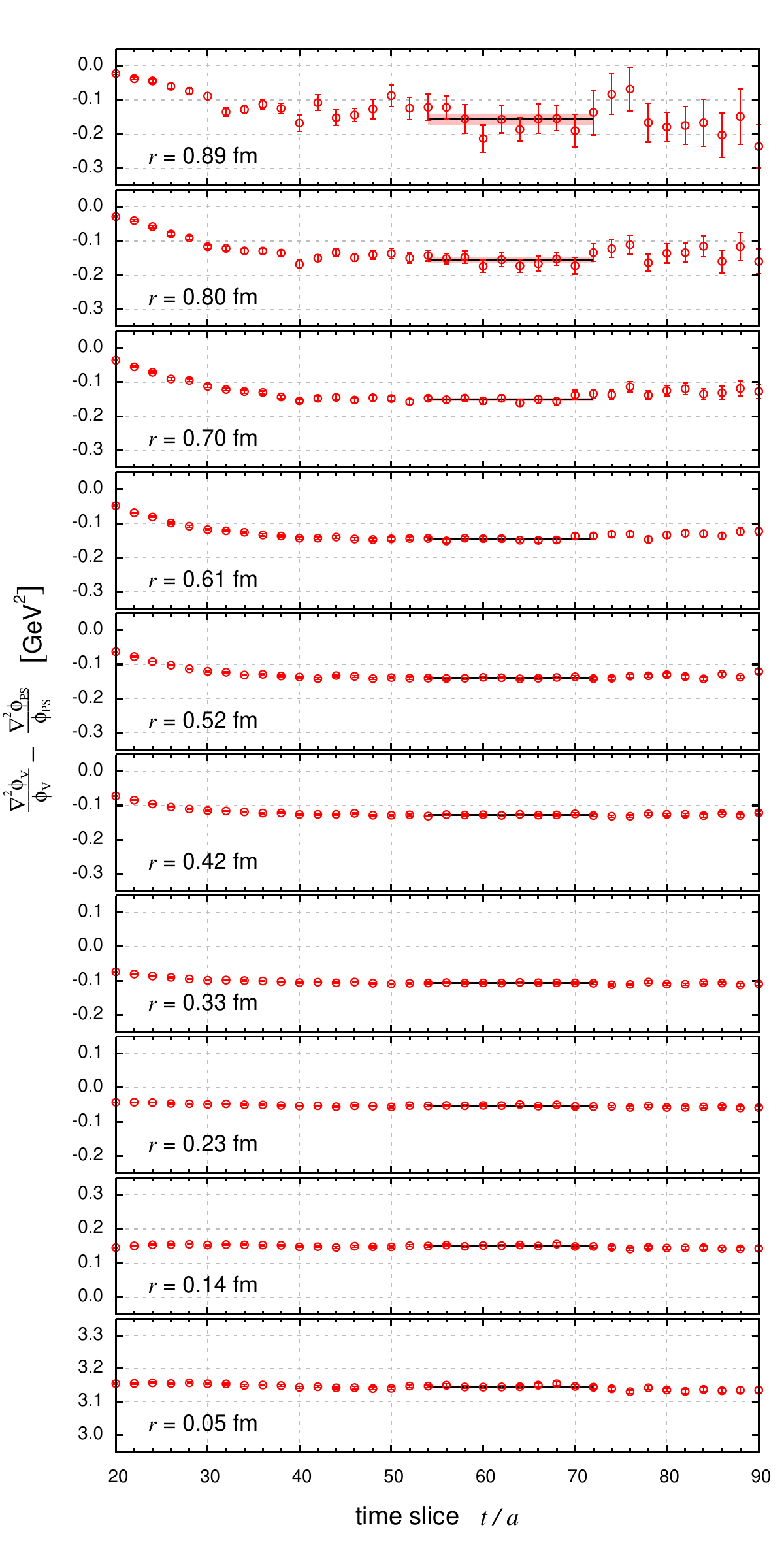}
  \caption{
    The time slice dependence of the spin-averaged ratio $\Vcal(r)$ (left)
    and the difference of ratios $\Vcal_S(r)$ (right)
    at various distances over the range $0.05~{\rm fm} \leq r \leq 0.89~{\rm fm}$,
    which are calculated on the FI ensembles at the charm quark mass as a typical example.
    \label{potential_tdep}
  }
 \end{figure*}
The ratios of $\nabla^2_{\rm lat}\phi_\Gamma/\phi_\Gamma$ at each spatial point $r$, shown in  Fig.~\ref{laplacian},
are actually  evaluated by a weighted sum of the corresponding ratios of
$\nabla^2_{\rm lat}G_\Gamma({\bf r}, t)/G_\Gamma({\bf r}, t)$ 
with respect to the time slice in the range, where the effective mass plot of the two-point function
shows the plateau. To resolve the strong correlations between data at different time slices, 
we take into account the full covariance matrix during the averaging process over the time slice.

Fig.~\ref{potential_tdep} shows 
time dependence of the the spin-averaged ratio 
$\Vcal(r) = \frac{3}{4}\nabla^2\phi_{\rm V}/\phi_{\rm V}+\frac{1}{4}\nabla^2\phi_{\rm PS}/\phi_{\rm PS}$
and the difference of ratios 
$\Vcal_S(r) = \nabla^2\phi_{\rm V}/\phi_{\rm V}-\nabla^2\phi_{\rm PS}/\phi_{\rm PS}$
calculated on the FI ensembles at charm quark mass as a typical example.
Both quantities are needed to calculate 
the spin-independent central, spin-spin potentials and quark kinetic mass through 
Eq.~(\ref{Eq_potC}), (\ref{Eq_potS}) and (\ref{eq_quark_mass}), respectively.
In Fig.~\ref{potential_tdep}, they exhibit reasonably long plateaus, and the asymptotic values 
at given $r$ can be read off from them.
Solid lines represent  average values over the plateau region. 
Shaded bands denote statistical errors estimated by the jackknife method.
There is no qualitative difference in the results obtained from the other ensembles (ME, CO and LA).

\subsection{\label{sec:determination_quarkmass}Quark kinetic mass}
\begin{table*}
\centering
\caption{Summary of the quark kinetic masses determined along three different directions
(on-axis, off-axis I and II) with the fit range $[r_\text{min}/\tilde{a}:r_\text{mas}/\tilde{a}]$
for all four ensembles.   
}
\label{tab:quarkmass}
\begin{ruledtabular}                                                                                                                  
 \begin{tabular}{c|ccc|ccc|ccc|c}
  &\multicolumn{3}{c|}{direction $(1,0,0)$, $\tilde{a} = a$}
  &\multicolumn{3}{c|}{direction $(1,1,0)$, $\tilde{a} = \sqrt{2}a$}
  &\multicolumn{3}{c|}{direction $(1,1,1)$, $\tilde{a} = \sqrt{3}a$}
  & average \\
  Label & fit range & $m_Q$[GeV] & $\chi^2/{\rm d.o.f.}$  
  & fit range & $m_Q$[GeV] & $\chi^2/{\rm d.o.f.}$  
  & fit range & $m_Q$[GeV] & $\chi^2/{\rm d.o.f.}$  
  & $m_Q$[GeV] \\ \hline
  FI & [14:20]& 1.982(56) & 1.11 & [10:14]& 1.997(52) & 1.68 & [8:11]& 2.030(50) & 1.62 & 2.013(43)\\
  ME & [9:14] & 1.967(50) & 0.60 & [7:10] & 1.990(60) & 0.44 & [6:8] & 1.984(73) & 0.34 & 1.980(55)\\
  CO & [7:10] & 1.937(39) & 0.63 & [5:7]  & 1.874(34) & 4.55 & [4:5] & 1.894(33) & 7.13 & 1.902(32)\\
  LA  & [7:13] & 1.874(39) & 0.86 &  [5:9] & 1.917(37) & 1.29 & [4:7] & 1.892(33) & 4.12 & 1.895(32) 
 \end{tabular}
\end{ruledtabular}                                                                                                                    
\end{table*}
 \begin{figure}
   \centering
   \includegraphics[width=.49\textwidth]{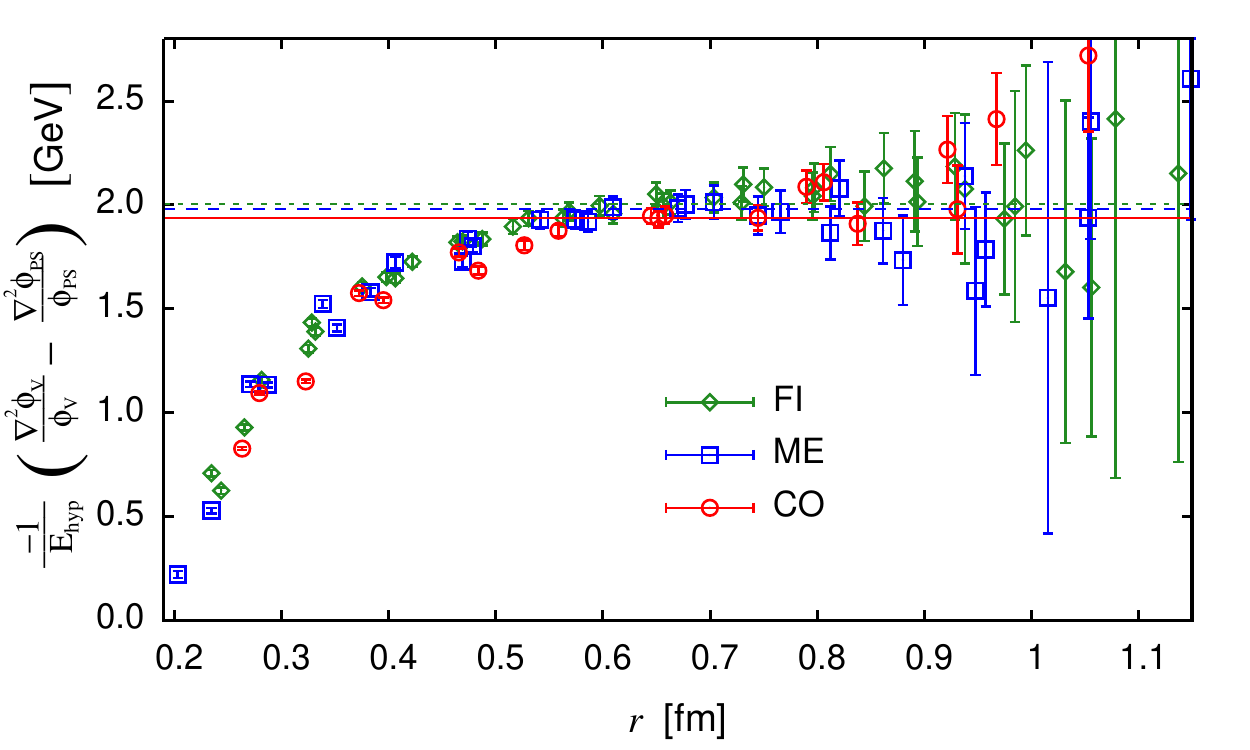}
   \caption{
    The determination of the quark kinetic mass within the BS amplitude method.
    The values of $-(\nabla^2\phi_{\rm V}/\phi_{\rm V}-\nabla^2\phi_{\rm PS}/\phi_{\rm PS})/E_\text{hyp}$
    as a function of the spatial distance $r$ are shown in this figure.
    Circle, square and diamond symbols denote results calculated on the FI, ME and CO ensembles, respectively.
    The quark kinetic masses $m_Q$ are evaluated from the long-distance asymptotic values of
    $-(\nabla^2\phi_{\rm V}/\phi_{\rm V}-\nabla^2\phi_{\rm PS}/\phi_{\rm PS})/E_\text{hyp}$.
    Horizontal solid (CO), dashed (ME) and dotted (FI) lines indicate results of the quark kinetic masses,
    which are determined by a weighted average of data points in the range $0.6~\text{fm}\alt r \alt 1.0~\text{fm}$
    as described in text.
  }
  \label{quarkmass}
 \end{figure}
 \begin{figure}
   \centering
   \includegraphics[width=.49\textwidth]{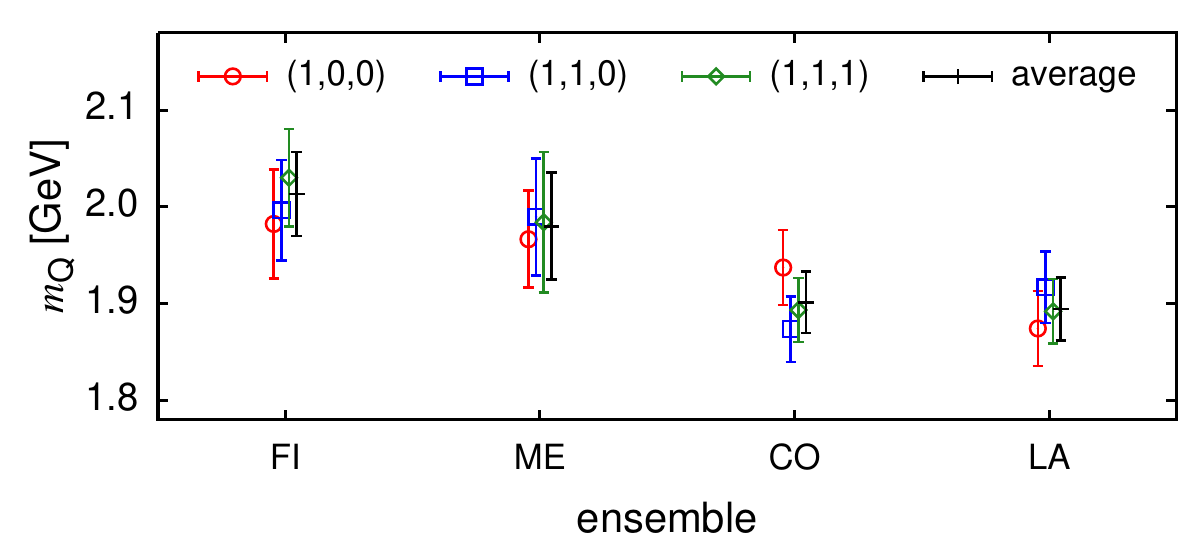}
   \caption{
    The quark kinetic mass calculated on all four ensembles.
    Circle, square and diamond symbols denote results
    calculated in the on-axis, off-axis~I and off-axis~II directions,
    respectively. Their averaged values are indicated by cross symbols.}
  \label{quarkmass_a}
 \end{figure}

In this subsection, we present the determination of the quark kinetic mass 
within the BS amplitude method.
A precise determination of the quark kinetic mass is required
for high-accuracy measurement of the interquark potentials.
In Fig.~\ref{quarkmass}, we plot the difference divided by the hyperfine splitting
energy at charm quark mass as a function of the spatial distance $r$.
At a glance, the value of $-(\nabla^2\phi_{\rm V}/\phi_{\rm V}-\nabla^2\phi_{\rm PS}/\phi_{\rm PS})/E_\text{hyp}$,
which appears in the r.h.s. of Eq.~(\ref{eq_quark_mass}),
certainly reaches a nonzero constant value at large distances and it 
turns out to be the value of the quark mass $m_Q$.

Practically the quark mass $m_Q$ is obtained by a constant fit to an asymptotic value 
over the range, where $V_{\rm S}(r)$ should vanish, taking into account the full covariance
matrix during the fitting process.
In this study, such constant fit is individually performed to the three data sets obtained from three directions:
on-axis, off-axis I and off-axis II. A difference of the quark masses obtained from
the different directions exposes the size of the possible finite size effect. We will quote it
as a systematic error on the quark kinetic mass. We finally take an average of the resulting masses over the three directions. 
The results of the quark kinetic mass are summarize in Table~\ref{tab:quarkmass} and also in
Fig.~\ref{quarkmass_a}.

As we mentioned in subsection \ref{sec:laplacian_op}, the discretization error introduced 
by the discrete Laplacian operator defined in Eq.~(\ref{eq_laplacian_polar})
along the off-axis~I and II directions are expected to be greater than that of the on-axis data.
Indeed we cannot obtain a reasonable value 
of $\chi^2/{\rm d.o.f.}$ from the constant fits onto the off-axis data for the CO and LA ensembles,
which are generated at the coarsest lattice spacing~(See Table~\ref{tab:quarkmass}). 
We also find that the lattice spacing dependence of the quark kinetic mass 
determined from the on-axis data is observed to be the smallest in Fig.~\ref{quarkmass_a}.
For the  CO and LA ensembles,
we therefore prefer to use the on-axis data solely in the analysis of the quark kinetic mass, 
instead of the averaged value over three directions, in the following discussion.

Final results on the quark kinetic mass calculated at the three
different lattice spacings (FI, ME and CO ensembles) show a good
agreement with each other. The largest difference among three results
is only less than a few \%.  Although our calibration of the RHQ parameters is not precise enough
as described in subsection~\ref{sec:effective_mass}, 
a good scaling is again observed in the quark kinetic mass within 
the current statistical precision. 

According to a direct comparison between results obtained 
in two different lattice volumes ($La\approx 3.0~{\rm fm}$ and $2.2~{\rm fm}$) 
at the coarsest lattice spacing (CO and LA ensembles),
the systematic uncertainty due to the finite volume effect is estimated as a few \% level. 
We confirm that there is no significant volume effect in our evaluation of the quark kinetic mass
even for the on-axis data.

\subsection{\label{sec:potential}Spin-independent interquark potential}
Using the quark mass determined in previous subsection, we can calculate both the spin-independent
central and spin-spin potential obtained from a set of the $\QQbar$ BS wave function $\phi_\Gamma(r)$ 
with $\Gamma=$ PS and V, through Eq.~(\ref{Eq_potC}) and (\ref{Eq_potS}).
The BS wave functions $\phi_\Gamma(r)$ are defined only by the ground state contributions of
the $r$-dependent amplitude $G_\Gamma(r,t)$.
We determine the values of interquark potentials $V(r)$ and $V_{\rm S}(r)$ 
by averaging over appropriate time-slice range (See subsection~\ref{sec:time_average}).

 \begin{figure}
   \centering
   \includegraphics[width=.49\textwidth]{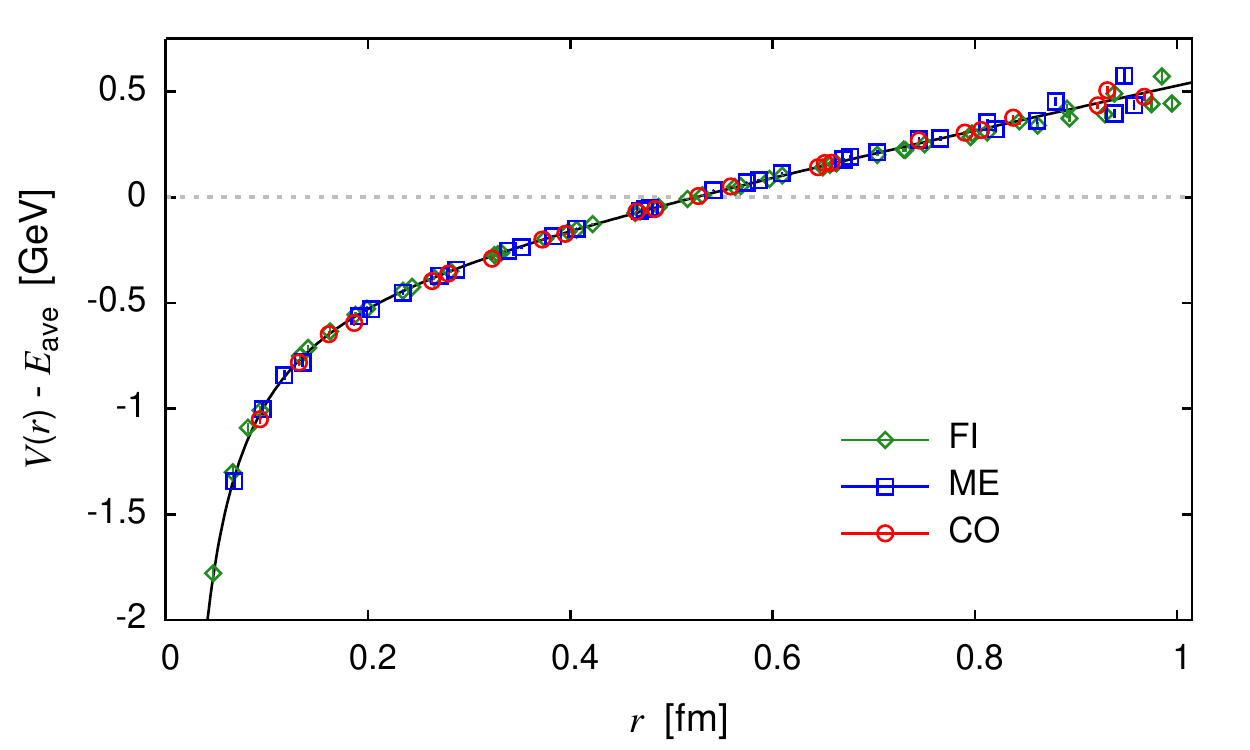}
   \includegraphics[width=.49\textwidth]{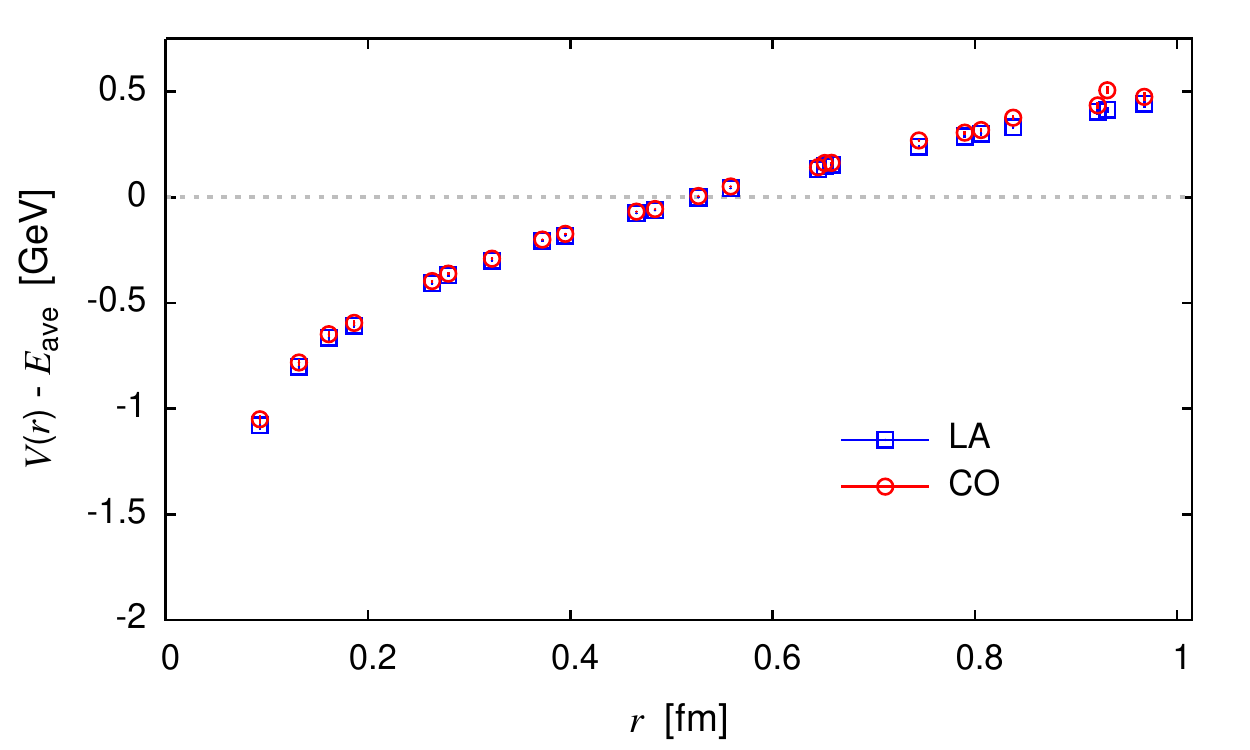}
  \caption{
    The lattice spacing dependence~(upper) and volume dependence~(lower) of
    spin-independent central potential~$V(r)$.
    For clarity of the figure, the constant energy shift $E_{\rm ave}$ is not subtracted.
    In the upper panel, a solid curve shows the fitting results of the Cornell potential form
    on the data points calculated on the FI ensembles. 
  }
  \label{potential}
 \end{figure}

The upper panel of Fig.~\ref{potential} shows all results of
 the spin-independent potential $V(r)$ at charm quark mass, 
that are calculated on three ensembles~({FI}, {ME} and {CO}) with fixed physical volume. 
For clarity of the figure, the constant energy shift $E_\text{ave}$, which 
corresponds to a value of $M_\text{ave}-2m_Q < 0$, is not subtracted in Fig.~\ref{potential}.
As expected, the resulting spin-independent central potential $V(r)$ with finite quark mass 
exhibits the linearly rising potential at large distances
and the Coulomb-like potential at short distances. 

In the upper panel of Fig.~\ref{potential},
the data points of the interquark potentials measured 
at different lattice spacings collapse on a single curve.
This would indicate that simulations at the gauge couplings $\beta =6/g^2= 6.0$, 6.2 and 6.47
are already in the asymptotic scaling region. Moreover we find the spin-independent central potential
determined from our proposed method can maintain the rotational symmetry accurately.

It is also worth noting that no adjustment parameter is added for showing a good scaling of
the interquark potential calculated at various $\beta$. 
This fact is contrast with the case of the static $\QQbar$ potential given by Wilson loops.
For the Wilson loop results, the constant self-energy contributions of infinitely heavy (static) 
color sources, which will diverge in the continuum limit, must be subtracted 
to demonstrate the scaling behavior.

The lower panel of Fig.~\ref{potential}
shows no visible finite volume effect on the spin-independent central potential $V(r)$
calculated at charm quark mass at least in the region of $r\alt 1$ fm. 
This observation is simply due to the fact that the $S$-wave BS wave function at charm quark mass
safely fits into even the smaller lattice volume ($La \approx 2.2$ fm).

\begin{table*}
\centering
\caption{Summary of the Cornell potential parameters ($A$, $\sqrt{\sigma}$ and unsubtracted $V_0$), 
  a ratio of $A/\sigma$ and the Sommer parameter~$r_0$, calculated on all four ensembles.
}
\label{tab:parameter_a}
\begin{ruledtabular}                                                                                                                  
 \begin{tabular}{clllll} 
  Label    & \multicolumn{1}{c}{$A$} & \multicolumn{1}{c}{$\sqrt{\sigma}$ [GeV]} & \multicolumn{1}{c}{$V_0 -E_{\rm ave}$ [GeV]} & \multicolumn{1}{c}{$A/\sigma$ [GeV$^{-2}$]}
  & \multicolumn{1}{c}{$r_0$ [rm]} \\ \hline
  FI    & 0.347(10)(28)(27)(15)&  0.439(7)(7)(12)(1)  & $-0.381$(15)(25)(37)(2) & 1.804(74)(207)(238)(66) & 0.512(8)(3)(8)(4)\\
  ME    & 0.390(13)(36)(25)(0) &  0.438(8)(5)(5)(3)   & $-0.356$(19)(26)(21)(7)  & 2.036(101)(239)(175)(27) & 0.505(10)(3)(1)(3) \\
  CO    & 0.382(10)(20)(10)(2) &  0.441(6)(5)(4)(3)   & $-0.370$(14)(26)(12)(5)  & 1.966(76)(132)(86)(15)   & 0.504(7)(2)(2)(4)\\  
  LA    & 0.442(11)(21)(27)(8) &  0.428(6)(11)(6)(5)  & $-0.324$(12)(29)(24)(7)  & 2.418(81)(175)(578)(9)   & 0.507(8)(13)(2)(8)\\
 \end{tabular}
\end{ruledtabular}                                                                                                                    
\end{table*}
\begin{figure}
   \centering
   \includegraphics[width=.49\textwidth]{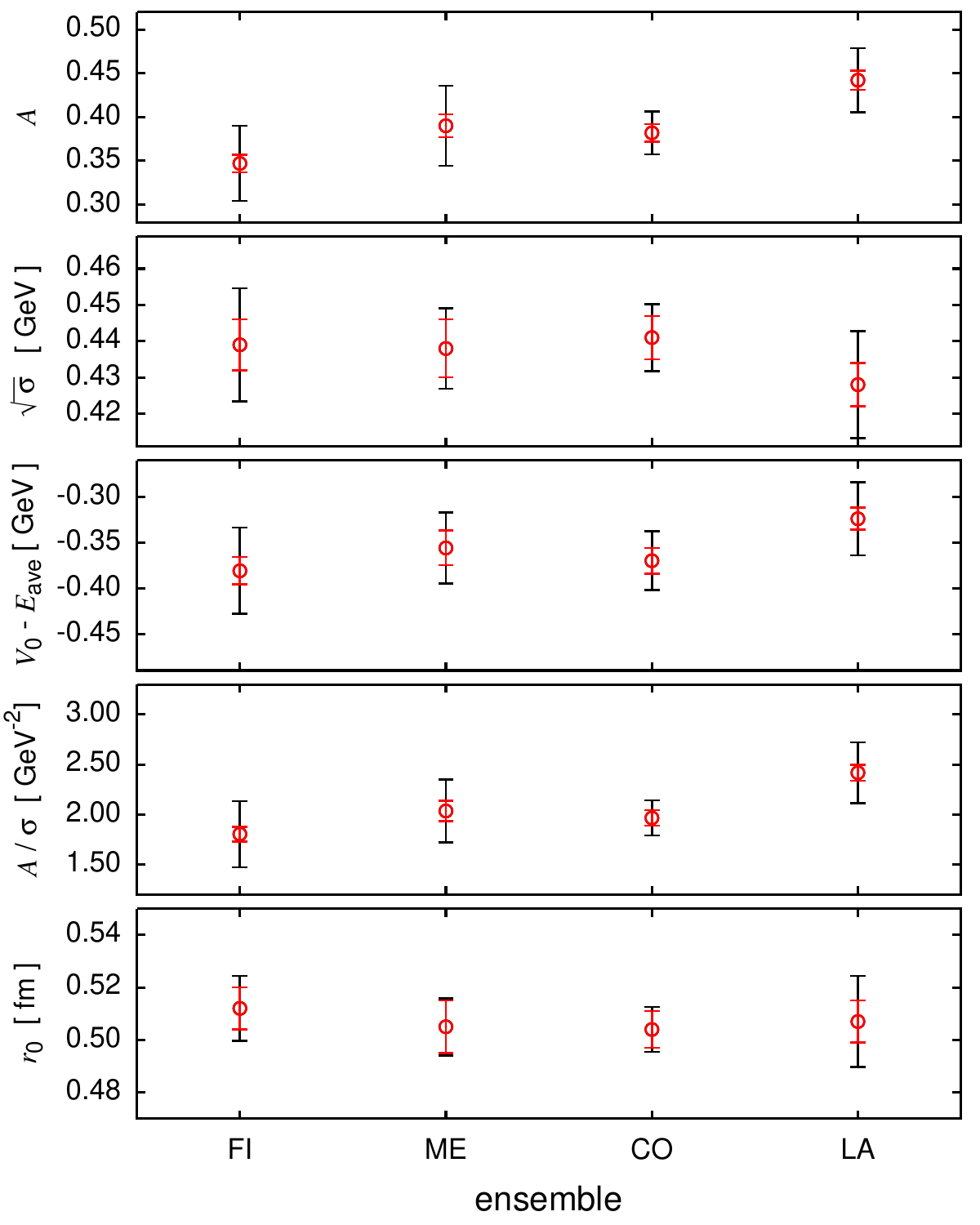}
   \caption{
    Summary of the Cornell potential parameters ($A$, $\sqrt{\sigma}$ and $V_0$), a ratio of $A/\sigma$ and the 
    Sommer scale~$r_0$ obtained from all four ensembles~({FI}, {ME}, {CO} and {LA}).
    The inner error bars indicate the statistical error,
    The outer error bars display the total errors, where the statistical and all systematic errors are added in quadrature. 
  }
  \label{parameters_a}
 \end{figure}

We simply adopt the Cornell potential parameterization for fitting the data of $V(r)$ as
\begin{equation}
 V(r)=-\frac{A}{r}+\sigma r+V_0  \label{eq_Cornell}
\end{equation}
with the Coulombic coefficient $A$, the string tension $\sigma$, and a constant $V_0$.
The Cornell potential parameterization describes well the spin-independent central potential even 
at finite quark mass.

Although the charm quark mass region would be beyond the radius of convergence 
for the systematic $1/\mQ$ expansion, 
the finite $\mQ$ corrections could be encoded into the Cornell potential parameters in this approach.
Table~\ref{tab:parameter_a} presents the summary of the Cornell potential parameters.
All fits are performed individually for the three directions (on-axis, off-axis I and II)
over the range $0.19~\text{fm}\alt r \alt 0.84~\text{fm}$.
We minimize the $\chi^2/{\rm d.o.f}$ taking into account the  covariance matrix.

In Table~\ref{tab:parameter_a}, all quoted values of the Cornell potential parameters are obtained 
by taking an average over the three directions.
The first errors are statistical ones. 
For the second errors, we estimate uncertainties of the choice 
of the data from the three direction and take the maximal 
difference from the average among the results of all three directions. 
Therefore the second errors are associated with the violation of the rotational symmetry.
The third and fourth ones are systematic uncertainties originating from the choice of 
minimum values ($t_{\rm min}$ and $r_{\rm min}$) of the temporal and spatial windows 
used in fitting procedures, respectively.

In addition, we estimate a ratio of $A/\sigma$ and the Sommer parameter~$r_0$, which are also included in Table~\ref{tab:parameter_a}.
The former is a quantity independent of the definition of the quark mass. 
In other words, it is simply related to a gross shape of the spin-independent 
central potential.
The later is a well-known phenomenological quantity defined by
\begin{equation}
 r_0^2 = \frac{d V(r)}{dr} \big{|}_{r=r_0} = 1.65.
\end{equation}
Thus, $r_0$ can be evaluated by the Cornell potential parameters as 
\begin{equation}
 r_0 = \sqrt{\frac{1.65 - A}{\sigma}}.
\end{equation}

Here we give a few technical remarks on the systematic uncertainties. 
The value of the string tension $\sigma$ is determined by 
the long-range behavior of the potential. However, the linear part
in the Cornell potential parameterization is dominated in the region
where we have data points. Thus, the resulting value of $\sigma$ 
is relatively insensitive to the choice of the fitting window ($r_{\rm min}, r_{\rm max}$)
and also the choice of the data set with respect to the direction, compared to 
the Coulombic coefficient $A$.  
A weak dependence of the latter suggests that a violation of the rotational symmetry 
is found to be small in the long-range part of the $\QQbar$ potential.
On the other hand, as we described above, the resulting
value of $A$ highly depends on the choice of the direction in the fitting procedure.
Therefore, there is a large systematic uncertainty associated with the rotational 
symmetry breaking. This indicates that the short-range part of the $\QQbar$ potential 
is not yet fully improved by reducing spatial discretization errors in the discrete Laplacian operator
as we proposed in subsection~\ref{sec:laplacian_op}. 

The fourth errors tabulated in Table~\ref{tab:parameter_a} are evaluated from 
uncertainties due to the choice of time window in the averaging process over the time slice.
These are the smallest errors among the other errors including the statistical one.  
This is attributed to the fact that we have taken a weighted average of data points
in the very wide range of time slices  as was discussed in subsection \ref{sec:time_average}.
This particularly contrasts with the conventional approach to calculate the static $\QQbar$ 
potential by Wilson-loops or Polyakov lines, where the largest systematic uncertainty
is due to the selection of their temporal length.

Fig.~\ref{parameters_a} displays the Cornell potential parameters ($A$, $\sqrt{\sigma}$, $V_0$),
a ratio of $A/\sigma$ and the Sommer scale~$r_0$,  obtained from all four ensembles~({FI}, {ME}, {CO} and {LA}), for comparison.
The inner and outer error bars are the statistical and total errors.
The total errors are given by the sum of statistical and systematic errors in quadrature. 
The resulting Cornell potential parameters calculated at various $\beta$ are consistent within their 
errors~(See results of the {FI}, {ME} and {CO} ensembles). On the other hand, 
although the results of the {CO} and {LA} ensembles are consistent within two
standard deviations, there appears to be a mild volume dependence on every parameter. 

It is worth mentioning here that $r_0$ is determined with high accuracy and 
has no obvious dependence of the lattice spacings and volumes.
Then $r_0$ agrees well with the input number of $r_0 = 0.5$~fm within errors.
This is attributed to the fact that the interquark potential at the range, 
where $V(r)-E_{\rm ave} \approx 0$, is most precisely determined in the BS amplitude method,
while $r_0$ is accidentally close to such region.   

\subsection{\label{sec:potential}Spin-Spin potential}
\begin{figure}
   \centering
   \includegraphics[width=.49\textwidth]{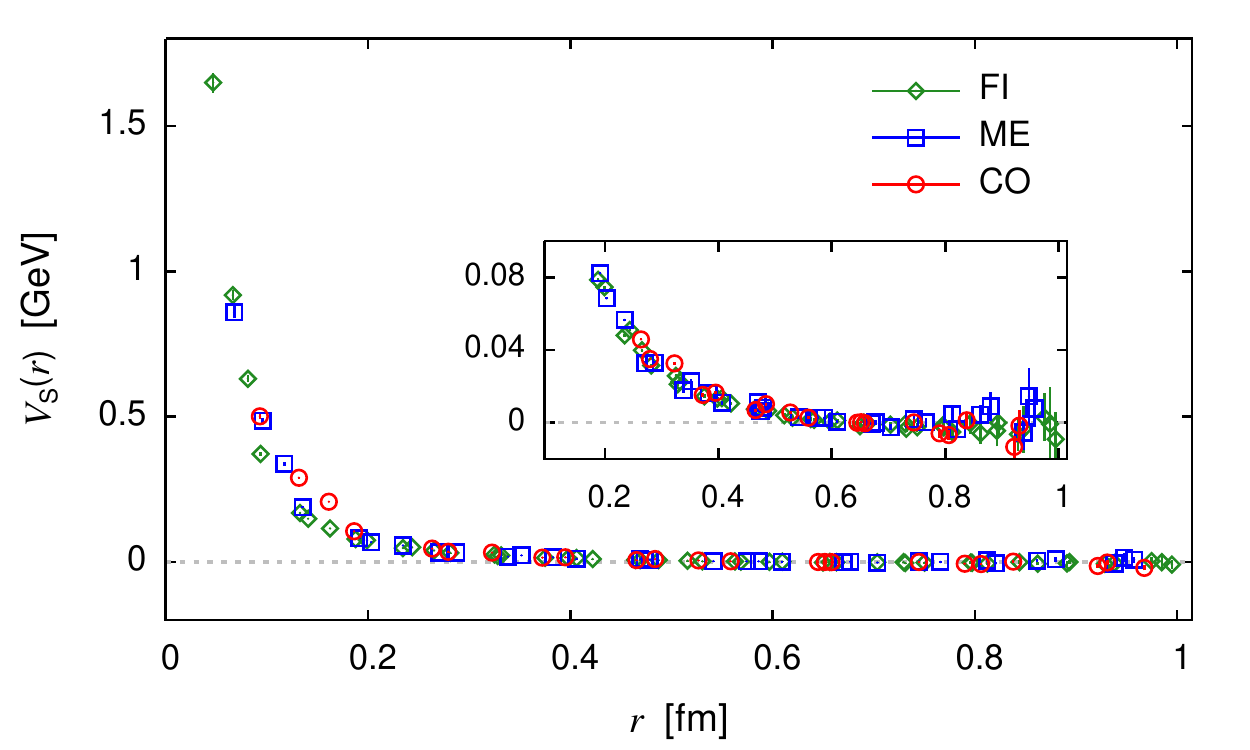}
   \includegraphics[width=.49\textwidth]{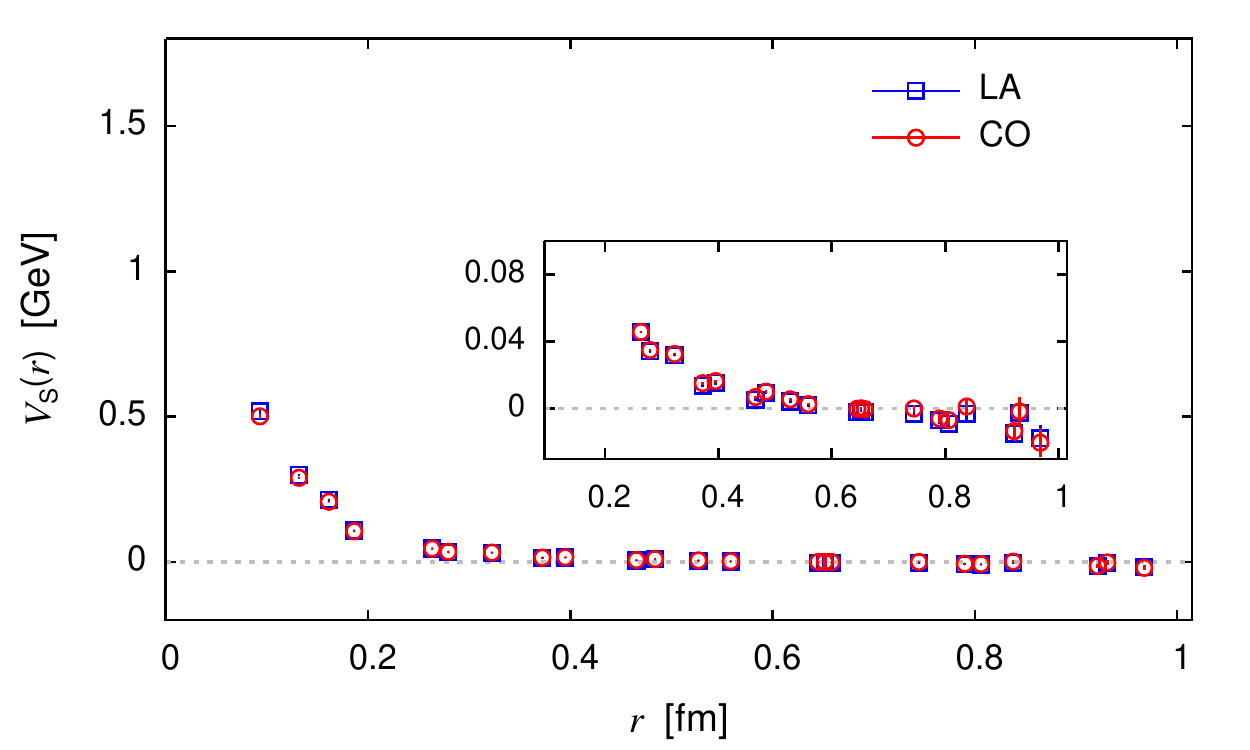}
   \caption{
    The lattice spacing dependence~(upper) and volume dependence~(lower) of
    the spin-spin potential~$V_{\rm S}(r)$.
    The insets show magnified views in the region of $r\agt 0.1~\text{fm}$.}
  \label{potential2}
 \end{figure}

We determine the spin-spin potential within the BS amplitude method,
through Eq.~(\ref{Eq_potS}),
similar to the spin-independent central potential $V(r)$.
Fig.~\ref{potential2} displays the spin-spin potential $V_{\rm S}(r)$
calculated from the $\QQbar$ BS wave function.
First the resulting potential is quickly dumped at large distances 
and exhibits a repulsive interaction with a finite range of $r \alt$ 0.6~fm. 
This is different from a short-range $\delta$-function potential based on 
one-gluon exchange like the Fermi-Breit interaction of QED.
Second repulsive interaction is required by the charmonium spectroscopy, 
where the higher spin state in hyperfine multiplets receives heavier mass. 
It should be reminded that the Wilson loop approach fails to reproduce 
the correct behavior of the spin-spin interaction even in the bottom sector.
The leading-order contribution of the spin-spin potential classified
in pNRQCD gives rise to a short-range attractive interaction,
which yields wrong mass ordering among hyperfine multiplets~\cite{Koma:2006fw}.  

As shown in the upper panel of Fig.~\ref{potential2},
the discretization artifacts are visible on the spin-spin potential at the short distances,
where the scaling behavior is violated. 
This contrasts to the spin-independent central potential, where a good scaling
behavior is observed even at the short distances.
However, this observation is consistent with the fact that 
the hyperfine splitting energies exhibit a slight, but systematic dependence of 
the lattice spacing~(See Fig.~\ref{hyperfine_E}).
In order to determine the spin-spin potential keeping systematics under control,
we will need simulations on finer lattices, or alternatively
perform nonperturbative tuning of the RHQ parameters and 
further improvement of the discrete Laplacian operator.

On the other hand, as for the finite volume effect, 
there is no significant difference between the spin-spin potentials
calculated from two different physical volumes (CO and LA) 
as shown in the lower panel of Fig.~\ref{potential2}.
This is consistent with the fact that the spin-spin potential $V_{\rm S}(r)$ is 
measured as the short-range potential and the BS wave function at short distances 
is insensitive to the spatial extent.

\section{\label{limit}Heavy quark mass limit of interquark potential}
%
%
\begin{table}
  \caption{
    Summary of the RHQ parameters ($\nu$, $r_s$, $c_B$ and  $c_E$) 
    and spin-averaged masses of the $1S$ heavy quarkonium state,
    used in the simulation with the FI ensembles toward the infinitely heavy quark limit.
    \label{RHQ_para_bottom}
      }
      \begin{ruledtabular}
      \begin{tabular}{ccccccl} 
	&$\kappa_Q$   &$\nu$ & $r_s$ & $c_B$ & $c_E$  &  
	\multicolumn{1}{c}{$M_{\rm ave}$ [GeV]}\\ \hline
	charm &0.11727  & 1.029 & 1.131 & 1.700 & 1.562 & 3.0676(20) \\
	&0.11198  & 1.041 & 1.165 & 1.749 & 1.581 & 3.9612(16) \\
	&0.10377  & 1.066 & 1.230 & 1.842 & 1.619 & 5.1925(13) \\
	&0.09004  & 1.124 & 1.364 & 2.033 & 1.708 & 7.2466(11) \\
	bottom&0.07619  & 1.211 & 1.543 & 2.388 & 1.839 & 9.4462(9) \\
	&0.05759  & 1.402 & 1.906 & 2.807 & 2.127 & 12.8013(8) \\
      \end{tabular}
      \end{ruledtabular}
\end{table}
In this section, we discuss an asymptotic behavior of 
both the spin-independent central  and spin-spin potentials 
in the heavy quark mass limit~$\mQ \to \infty$.
We will first show that the spin-independent central potential in the $\mQ \to \infty$ limit is
fairly consistent with the conventional one obtained from Wilson-loops or Polyakov lines.
For this purpose,  we examine the quark mass dependence 
of the potentials near the infinitely heavy quark mass as much as possible. 

To avoid further discretization errors induced by heavier quark masses,
 we choose the finest lattice spacing ensembles (FI) and perform additional simulations 
with extra five hopping parameters, which corresponds to the heavier quark masses than the charm quark sector.  
The inverse of lattice spacing on the FI ensembles is about 4.2 GeV, which is
closet to the bottom mass.  Therefore, we choose our hopping parameters 
covering a wide mass range from the charm to beyond the bottom region 
toward the heavy quark limit. 

At the second heaviest quark mass
($\kappa_Q = 0.07619$), we obtain the spin-averaged $1S$-heavy quarkonium mass as $M_{\rm ave} = 9.4462(9)~{\rm GeV}$, which is close to the experimental one of the bottomonium.
Thus, $\kappa_Q = 0.07619$ is reserved for the bottom quark mass. 
It is worth mentioning that the hyperfine splitting energy calculated at the bottom quark mass 
in our simulations reproduces only 40\% of the experimental value~\cite{Beringer:1900zz}.
At each $\kappa_Q$, we again use the one-loop perturbation theory to determine five RHQ parameters following Ref.~\cite{Kayaba:2006cg}.
These RHQ parameters, which are summarized with given values of $\kappa_Q$ 
in Table~\ref{RHQ_para_bottom}, marginally satisfies the condition of
$c_{\rm eff}^2 = 1$ for the $1S$ heavy quarkonium states at all five quark masses
within errors. 

\subsection{BS wave function}
\begin{figure}
   \centering
   \includegraphics[width=.49\textwidth]{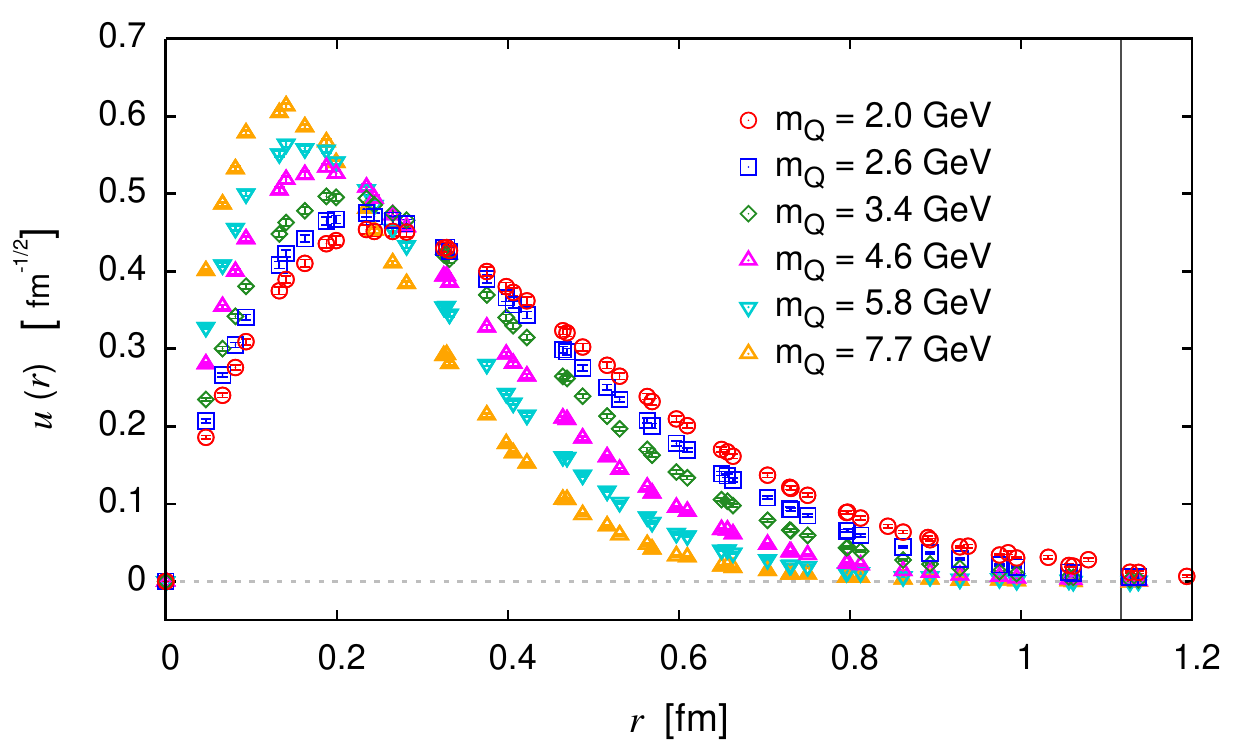}
   \caption{
    The reduced $\QQbar$ BS wave functions of the pseudoscalar quarkonium state calculated 
    using the FI ensembles with six different quark masses covering the range 
    from 2.0~GeV to 7.7~GeV, shown as a function of the spatial distance of~$r$.
    A vertical solid line denotes a half of the lattice size~($La/2\approx 1.1$~fm).
}
  \label{wavefunction_to_bottom}
 \end{figure}
In Fig.~\ref{wavefunction_to_bottom}, we first plot the reduced $\QQbar$ BS wave functions 
of the pseudoscalar quarkonium calculated at various quark masses. 
These wave functions are normalized as to fulfill the condition $\sum \tilde{\phi}^2 = 1$.
We again find the isotropic behavior in the BS wave functions even
at around the bottom quark mass.
The data points calculated from the three 
directions basically collapse on a single curve.
Nothing changes for the vector quarkonium wave function.

The wave function with a heavier quark mass is more localized than the one with a lighter quark mass.
Thus, the finite volume effect on the interquark potential becomes not serious at around the bottom quark mass. 
For the price one has to pay, a number of accessible data points at long distances gradually reduces for heavier quark mass.
It is worth reminding that in the BS amplitude method, we cannot access the information of the interquark potential outside of the localized wave function, where the wave function  
approximately vanishes and a signal-to-noise ratio in $\nabla^2\phi_\Gamma/\phi_\Gamma$ gets worse.

\subsection{Spin-independent interquark potential}
\begin{figure}
   \centering
   \includegraphics[width=.49\textwidth]{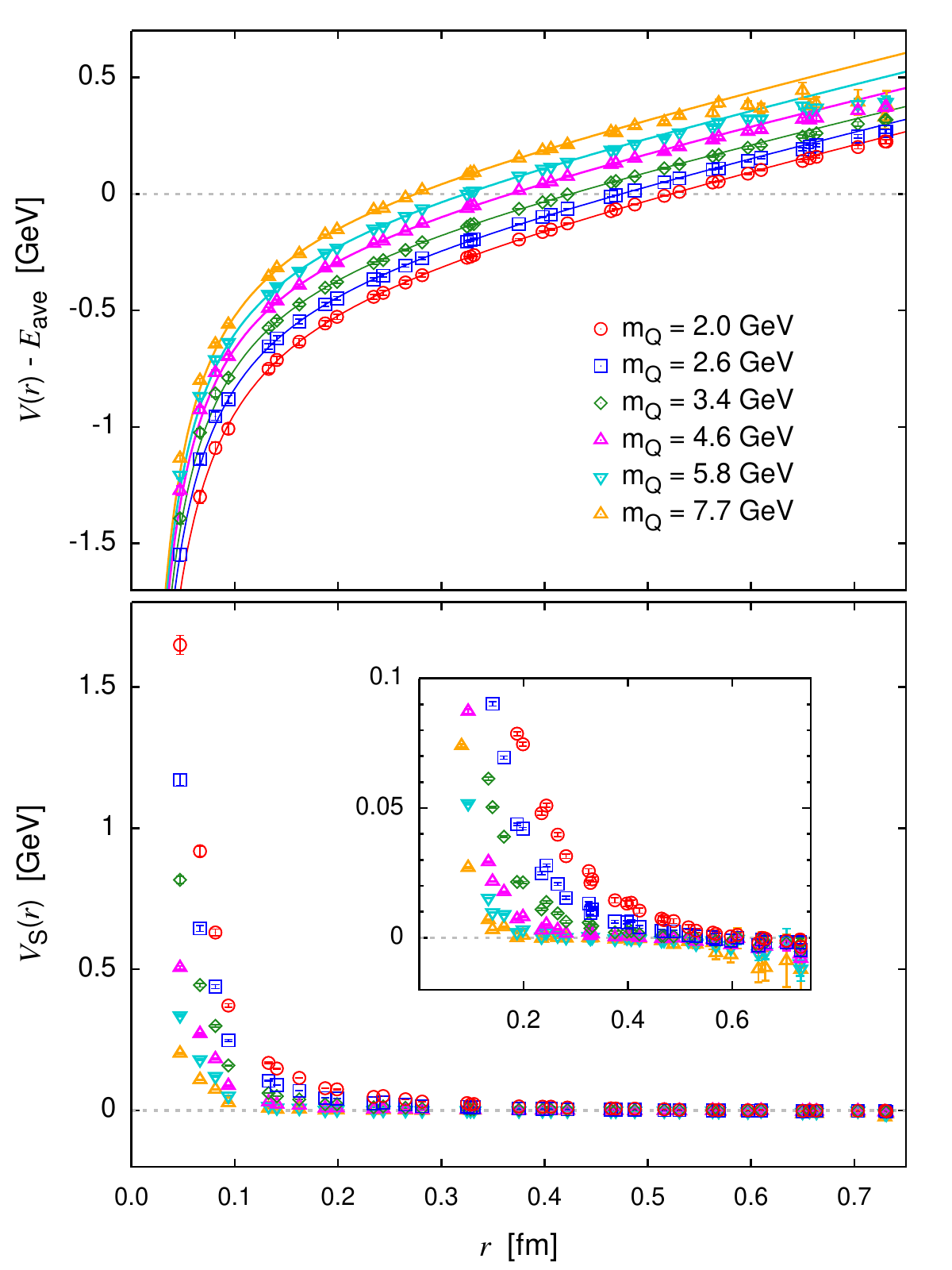}
   \caption{
    The spin-independent central (upper) and spin-spin (lower) potential calculated from the $\QQbar$ BS wave function
    at finite quark masses covering the range from 2.0~GeV to 7.7~GeV. 
    In the upper panel, each curve represents the fitting result of the Cornell potential 
    form given in Eq.~(\ref{eq_Cornell}), and also the constant energy shift $E_{\rm ave}$ is not subtracted.
    The inset in the lower panel shows a magnified view in the region of $r\alt 0.6~\text{fm}$.
}
  \label{potential_to_bottom}
\end{figure}
%

%
%
\begin{table}
  \caption{    
    Results of the quark kinetic mass $\mQ$, 
    the Cornell potential parameters $A$, $\sqrt{\sigma}$, and the ratio $A/\sigma$,
    calculated on the FI ensembles.
    Their extrapolated values in the $\mQ \to \infty$ limit using linear and quadratic fit forms are 
    compared with our results given by the Polyakov line correlator and also accurate results 
    calculated with the multilevel algorithm~\cite{Koma:2006fw}.~\label{cornell_bottom}
      }
      \begin{ruledtabular}
      \begin{tabular}{lcllc} 
	  \multicolumn{1}{l}{$\kappa_Q$}  & $\mQ$& \multicolumn{1}{c}{$A$} & \multicolumn{1}{c}{$\sqrt{\sigma}$} & $A/\sigma$\\
	    & [GeV] & &\multicolumn{1}{c}{[GeV]}&[${\rm GeV}^{-2}$] \\
 \hline
	0.11727 & 2.00(5) & 0.323(9) & 0.447(6) & 1.62(5) \\ 
	0.11198 & 2.60(5) & 0.297(6) & 0.443(5) & 1.51(4) \\ 
	0.10377 & 3.36(6) & 0.288(6) & 0.439(5) & 1.49(5) \\ 
	0.09004 & 4.57(7) & 0.279(5) & 0.441(5) & 1.43(4) \\ 
	0.07619 & 5.80(7) & 0.277(4) & 0.445(5) & 1.40(4) \\ 
	0.05759 & 7.71(8) & 0.277(4) & 0.446(5) & 1.39(4) \\
        \hline
	linear fit   & $\infty$ &0.273(9)  &0.454(11) &1.31(9) \\
	quadratic fit& $\infty$ &0.285(11) &0.454(12) &1.40(9) \\
	\hline
	\multicolumn{2}{l}{static $\QQbar$ (Polyakov lines)}   & 0.285(11)& 0.467(6) & 1.31(8) \\
	\multicolumn{2}{l}{static $\QQbar$ (Ref.~\cite{Koma:2006fw})}   & 0.281(5) & 0.458(1) &1.34(2) 
      \end{tabular}
      \end{ruledtabular}
\end{table}

Fig.~\ref{potential_to_bottom} displays 
the spin-independent central potential (upper) and 
spin-spin potential (lower) calculated at several quark masses
within the BS amplitude method.
In the upper panel of Fig.~\ref{potential_to_bottom}, 
the constant energy shift $E_{\rm ave}$ is not subtracted as same in Fig.~\ref{potential}.
At first glance, the ``Coulomb plus confining potential'' are
observed over range from the charm to the bottom quark mass.
We perform a fit of the potentials calculated at various quark masses
to a simple form of the Coulomb plus linear potential, then obtain the Cornell potential parameters,
which are summarized in Table.~\ref{cornell_bottom}.
All fits are performed over the range $3\leq r/a \leq 7\sqrt{3}$  by correlated $\chi^2$ fit.
The errors quoted in Table.~\ref{cornell_bottom} are only statistical uncertainties, which are  
estimated by the jackknife method.

\begin{figure}
   \centering
   \includegraphics[width=.49\textwidth]{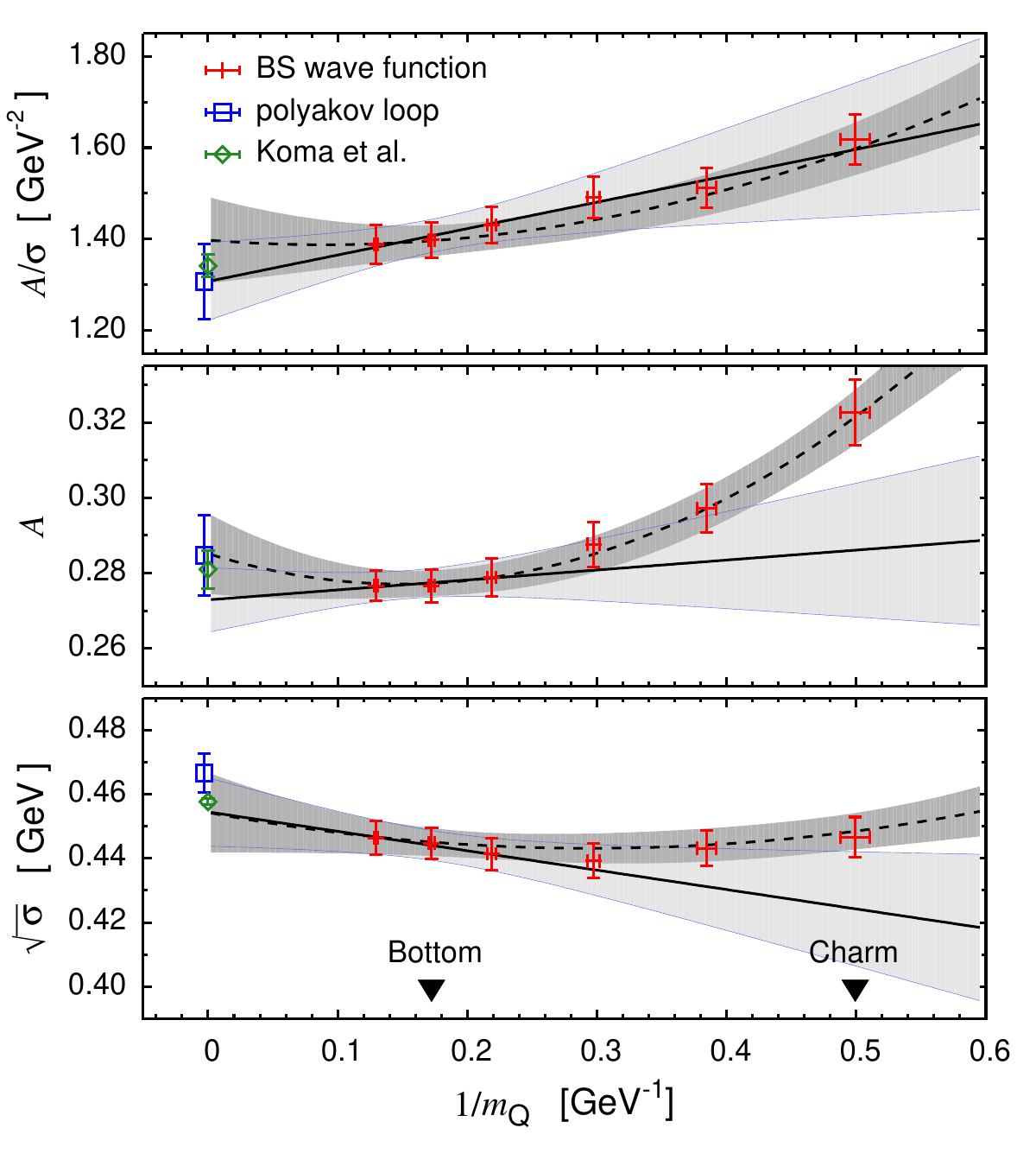}
   \caption{The quark-mass dependence of $A/\sigma$ (upper), $A$ (middle) and $\sqrt{\sigma}$ (lower),
    shown as functions of $1/\mQ$. 
    We perform the extrapolation toward the $\mQ \to \infty$ limit of 
    $A/\sigma$, $A$ and $\sigma$ with a simple polynomial function in $1/\mQ$.        
    Solid lines and dashed curves in each panel indicate the fitting results of linear and quadratic 
    forms, respectively.
    Shaded bands show statistical fitting uncertainties estimated by the jackknife method.
    The results of the static $\QQbar$ potential calculated by the Polyakov line correlator and also the Wilson loop
     using the multilevel algorithm~\cite{Koma:2006fw}   
    are also included as square and diamond symbols.
  }
  \label{parameters}
 \end{figure}
In Fig.~\ref{parameters}, we show the quark-mass dependence of the ratio of $A/\sigma$ (upper), 
the Coulombic coefficient $A$ (middle) and the squared-string tension $\sqrt{\sigma}$ (lower).
We also include values of the static $\QQbar$ potential calculated from the Polyakov line
correlator $P(r,t)$ as reference values in the infinitely heavy quark limit.
The static $\QQbar$ potential are obtained by fitting a plateau of the effective potential
$ V_{\rm eff}(r,t) = \ln \left\{{P(r,t)}/{P(r,t+1)}\right\}$
over range $[t_{\rm min}, t_{\rm mas}] = [7:10]$.
The Cornell potential parameters can be obtained by applying the same fitting procedure 
used in the case of the BS amplitude method.
We additionally include more accurate results given by Wilson loops using the multilevel algorithm~\cite{Koma:2006fw}.

First, regardless of the definition of $\mQ$, the ratio of $A/\sigma$ 
in the upper panel of Fig.~\ref{parameters} indicates 
that the interquark potential calculated from the BS wave function smoothly approaches 
the one obtained from Wilson loops in the infinitely heavy quark limit. 
The extrapolation toward the $\mQ\rightarrow \infty$ limit
is consistent with the value obtained from the static $\QQbar$ potentials.
Here, we perform both linear~(solid line) and  quadratic~(dashed curve) fits
with respect to $1/\mQ$ to three heaviest points and all of six data points, respectively.
All fits take into account the correlations among the different mass data in correlated $\chi^2$ fit. 
Shaded bands appeared in Fig.~\ref{parameters} indicate 
statistical errors, which are estimated by the jackknife method.

Second, if we pay attention to the quark-mass dependence of each of the Cornell potential parameters separately,
we observe that the Coulombic parameter $A$ depends on the quark mass significantly, while
there is no appreciable dependence of the quark mass on the string tension 
(see the middle and lower panels in Fig.~\ref{parameters}).
The finite $m_Q$ corrections seem to appear mainly in  
the short-range part of the potential characterized by the Coulombic coefficient~$A$.
At the charm quark mass, higher order corrections, at least the $\Ocal(1/m_Q^2)$ corrections, 
could be quite important to describe the spin-independent central potential.

We finally evaluate the values of $A$ and $\sqrt{\sigma}$ in the infinitely heavy quark limit
by both quadratic and linear fits 
as shown in Fig.~\ref{parameters}, and also the results are summarized in Table~\ref{cornell_bottom}.
Extrapolated values in the $\mQ \to \infty$ limit are consistent with those of the static $\QQbar$ 
potentials. We stress that our proposed method for determining the interquark potential 
with the proper quark mass given in Eq.~(\ref{eq_quark_mass}) is responsible for 
the quark-mass dependence observed here.

\subsection{Spin-spin potential}
\begin{figure}
   \centering
   \includegraphics[width=.49\textwidth]{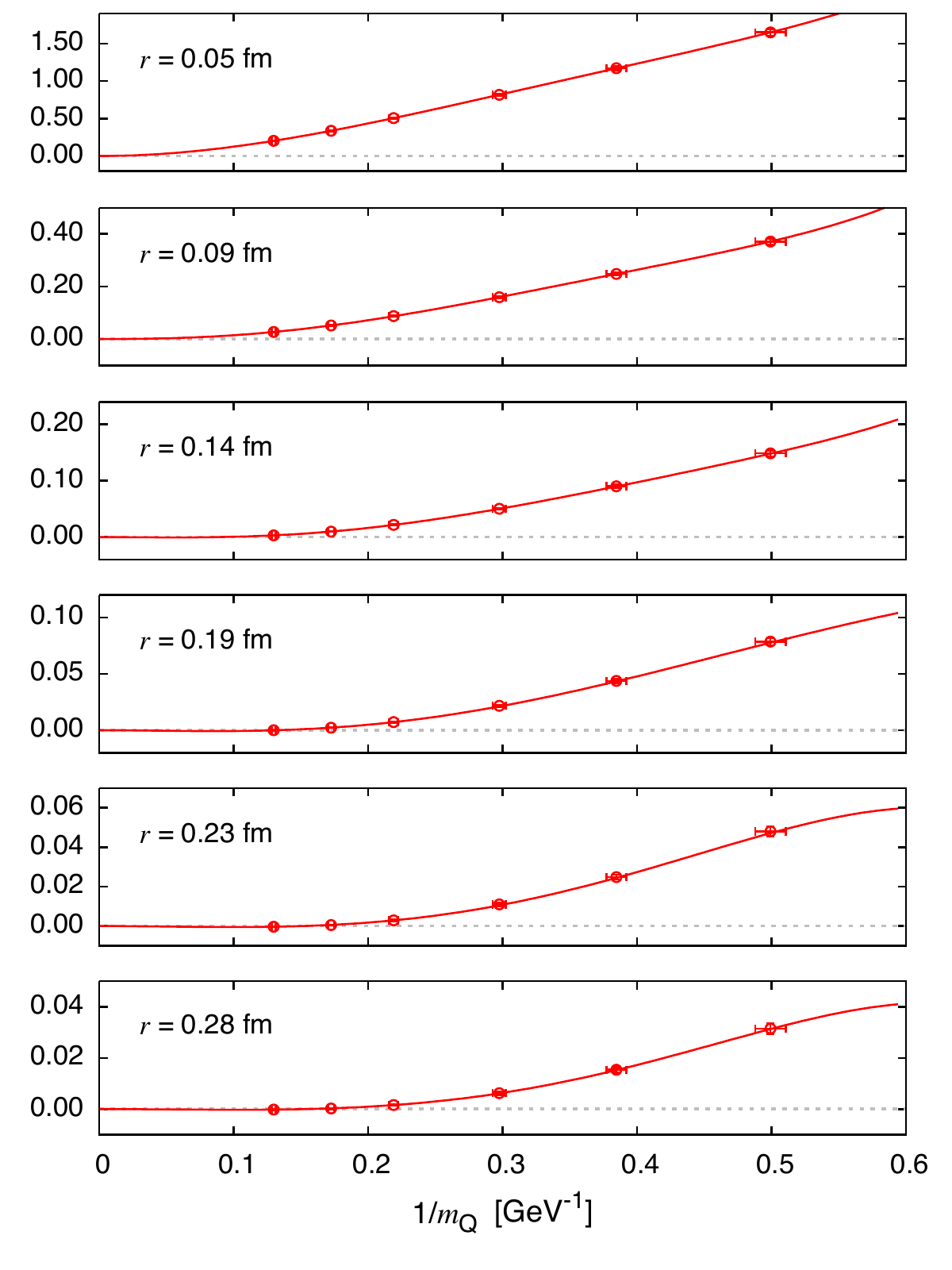}
   \caption{    
    The quark mass dependences of the spin-spin potential~$V_{\rm S}(r)$ at fixed $r$ as functions
    of $1/m_{\mQ}$.
    The selected values of $r$ are indicated in each panel. 
    The vertical axis is plotted in units of GeV.
    Solid curves correspond to fitting results of the polynomial
    forms given in Eq.~(\ref{eq_massdep_spin}).
  }
  \label{spin_potential_fit}
 \end{figure}
The quark-mass dependence of the spin-spin potential is more pronounced 
in contrast to the spin-independent central potential~(see the lower panel of Fig.~\ref{potential_to_bottom}). 
As the quark mass increases, a finite range of the spin-spin interaction becomes narrower, 
and then the potential seems to approach the $\delta$-function potential,
which would be induced by one-gluon exchange. 
We may expect that the spin-spin potential obtained in the BS amplitude method has
a correct behavior toward the $m_Q \rightarrow \infty$ limit. 

The spin-dependent potential in pNRQCD appears  as the $1/\mQ$ corrections 
 to the static $\QQbar$ potential. 
However there is a huge gap between our spin-spin potential at finite quark mass 
and  one  determined at ${\cal O}(1/\mQ^2)$ within 
the systematic $1/m_Q$ expansion approach~\cite{Koma:2006fw,Koma:2010zz}.
The former exhibits the short-range {\it repulsive} interaction, while 
the latter is similarly short-ranged, but turns out to be {\it slight attractive} 
interaction near the origin. 

To resolve the issue of the qualitative difference between two methods,
we try to read off the corresponding leading and also higher order corrections in the $1/\mQ$ expansion
 from our spin-spin potential, where all orders in the $1/m_Q$ expansion 
are supposed to be non-perturbatively encoded. 
We thus try to parametrize the spin-spin potential calculated with the finite quark mass
$m_Q$ in guidance of pNRQCD~\footnote{
Odd powers of $1/\mQ$ could appear in the case of non-abelian gauge theory~\cite{YS}.} as
\begin{equation}
 V_{\rm S} (\mQ, r) =  \frac{1}{\mQ^2}\left( V_{\rm S}^{(2)}(r)   + \frac{1}{\mQ} V_{\rm S}^{(3)}(r)  + \cdots
 \right).
\label{eq_massdep_spin}
\end{equation}
In Refs.~~\cite{Koma:2006fw,Koma:2010zz}, 
the leading order contribution of $V_{\rm S}^{(2)}(r)$
is precisely determined within the Wilson loop formalism 
using the multilevel algorithm. As was already mentioned, 
their spin-spin potential exhibits slight attractive interaction near the origin.

In Fig.~\ref{spin_potential_fit} we plot the spin-spin potential at fixed $r$ as a function of $1/m_Q$. 
At every $r$, we have carried out correlated $\chi^2$ fits on all six data displayed in Fig.~\ref{spin_potential_fit} 
by using a polynomial form of $1/m_Q$, according to Eq.~(\ref{eq_massdep_spin}). 
The $m$-th coefficient of the polynomial expansion with respect to 
$1/m_Q$ can be identified as the potential value of $V_{\rm S}^{(m+1)}(r)$ at given $r$, 
corresponding to the correction term at $\Ocal(1/\mQ^{m+1})$~\footnote{
The same analysis, in principle, can be applied to the spin-independent central potential.
The leading order potential $V^{(0)}(r)$, which corresponds to the $\QQbar$ potential
in the $\mQ \rightarrow \infty$ limit, was obtained in this procedure. We have confirmed
that $V^{(0)}(r)$ obtained in this analysis is fairly consistent with the static $\QQbar$ potential 
calculated from the Polyakov line correlator.  
However, the spin-independent central potential involves the self energy of a quark and
anti-quark pair, which is proportional to $m_Q$ as 
\begin{equation}
V (\mQ, r) =  {\rm constant} \times m_Q + V^{(0)}(r) +  \frac{1}{\mQ} V_{\rm S}^{(1)}(r)  + \cdots. 
\end{equation}
The presence of a term of ${\cal O}(\mQ)$ in addition to the polynomial function of $1/\mQ$ 
makes the fit relatively unstable, compared to the case of the spin-spin potential. 
Unfortunately, we did not observe the stability of the fit results even for 
the leading order correction of $\Ocal(1/{\mQ})$ within the current statistics. 
}.
The fit results are also displayed as solid curves in Fig.~\ref{spin_potential_fit}.
The stability of the fit results has been tested against either the number of 
fitted data points or the number of the polynomial terms.
We find that the polynomial terms up to the $\Ocal(1/\mQ^{5})$ term
are necessary to describe the quark mass dependence of the spin-spin potential,
covering a whole range of 2.0~GeV $\leq\mQ\leq$ 7.7~GeV, due to the 
slow convergence of the $1/\mQ$ expansion in the vicinity of the charm sector.
Our choice of the maximum polynomial term of ${\cal O}(1/\mQ^5)$ 
in the fitting form as Eq.~(\ref{eq_massdep_spin}) certainly yields acceptable 
values of $\chi^2/{\rm d.o.f}$ and confidence level. 

\begin{figure}
   \centering
   \includegraphics[width=.49\textwidth]{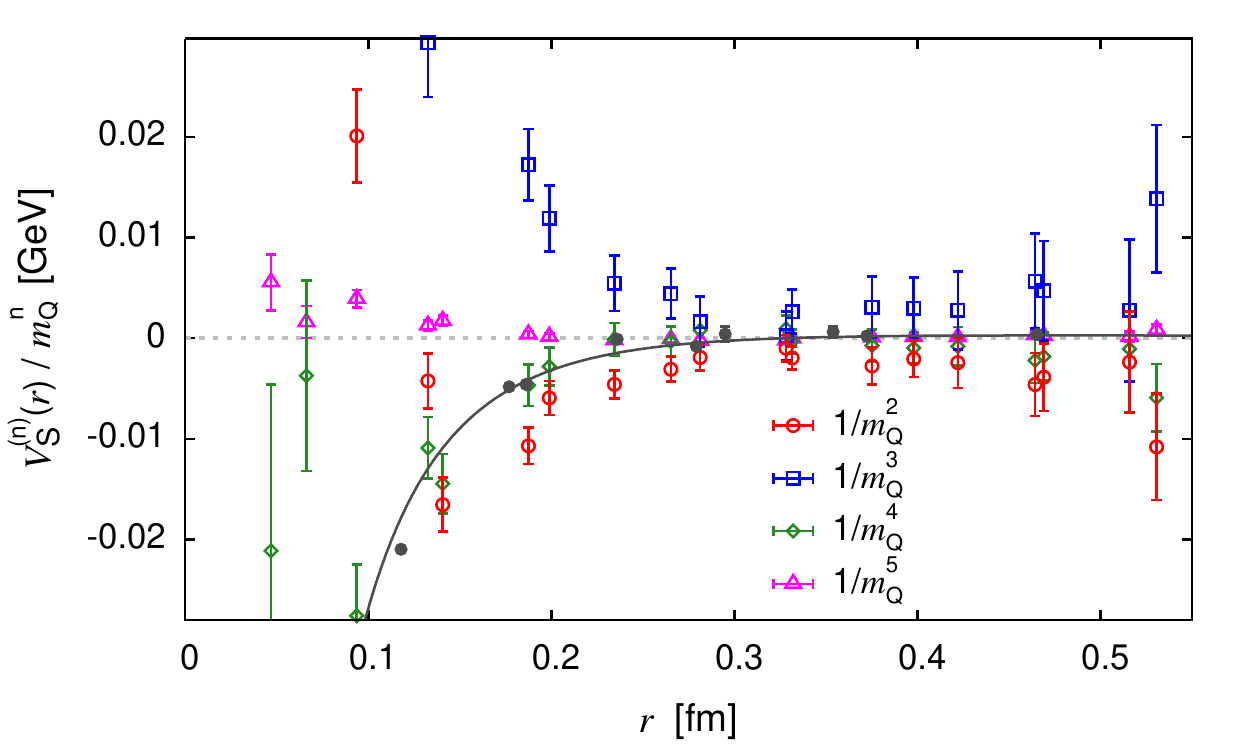}
   \includegraphics[width=.49\textwidth]{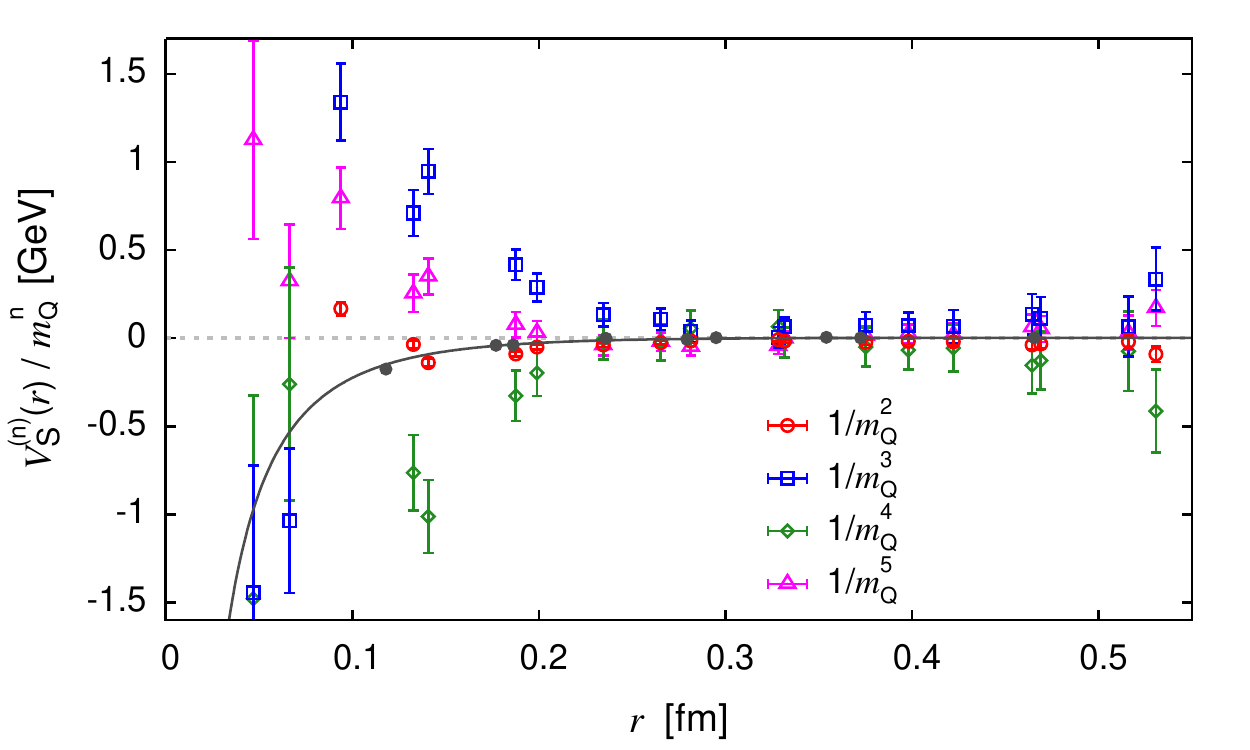}
   \caption{
    The $r$-dependence of the inverse quark mass corrections $V_{\rm S}^{(n)}(r)/\mQ^n$ 
    on the spin-spin potential at the bottom~(upper) and charm~(lower) quark masses.
    Filled circles correspond to the spin-spin potential at $\Ocal(1/\mQ^2)$ calculated within the Wilson 
    loop formalism, together with their fit results (solid curves)~\cite{Koma:2006fw,Koma:2010zz}.
  }
  \label{m_corre_spin_potential}
 \end{figure}

In~Fig.~\ref{m_corre_spin_potential}, we compile all results of $V_{\rm S}^{(n)}(r)$
(up to $n=5$), scaling with powers of $1/\mQ^n$, in order to analyze
the convergence behavior of the $1/\mQ$ expansion at both the bottom (upper)
and charm (lower) quark masses. 
As shown in the upper panel of Fig.~\ref{m_corre_spin_potential}, 
the $\Ocal({1/\mQ^2})$ contribution (open circles) to the total spin-spin potential 
exhibits an exponentially screened at the long distances and attractive interaction 
in the intermediate region ($0.1~\text{fm}\alt r \alt 0.3~\text{fm}$).
Surprisingly, the $\Ocal({1/\mQ^3})$ contribution (open squares) is
the largest contribution, and ensures the short-range repulsive interaction 
of the total spin-spin potential.

Here, we remark on the short-range behavior found 
in the $\Ocal({1/\mQ^2})$ contribution near the origin. At the short 
distances ($r\alt 0.1~\text{fm}$), the sign of the spin-spin potential changes from 
negative to positive. 
We will later explain the reason why we do not take it seriously and then let us
focus on results obtained in the region of $r\agt 0.1~\text{fm}$.

The solid curve represents the fit curve on the data points (filled circles) taken 
from Ref.~\cite{Koma:2006fw,Koma:2010zz}, scaled by $1/\mQ^2$ with the bottom 
quark mass, $\mQ=5.80(7)$ GeV. The size of the attraction found in the $\Ocal({1/\mQ^2})$ 
contribution is almost the same order of magnitude as
that of the spin-spin potential determined within the Wilson loop formalism~\cite{Koma:2006fw,Koma:2010zz}. 

At this point, we may have a hint to fill out a gap between our results of the spin-spin potential 
calculated in the BS amplitude method and one calculated at $\Ocal({1/\mQ^2})$ 
within the $1/\mQ$ expansion scheme. According our analysis, the next-to-leading
order contribution of $\Ocal({1/\mQ^3})$ is not negligible, rather a dominant contribution
in the full spin-spin potential. In other words, the issue of the spin-spin potential in the $1/\mQ$ 
expansion approach within the Wilson loop formalism would be cured by the next-leading-order
contribution of $\Ocal({1/\mQ^3})$. 
Furthermore, although the sizes of $\Ocal({1/\mQ^2})$ and $\Ocal({1/\mQ^3})$ contributions are  
inverted in the sense of the systematic $1/\mQ$ expansion, 
the higher order contributions are certainly smaller than a sum of the two lowest contributions at the bottom quark mass.
Therefore, our analysis suggests that the $1/\mQ$ expansion scheme may 
have the convergence behavior up to the bottom sector. 

It is, however, not the case for the charm sector. 
In the lower panel of Fig.~\ref{m_corre_spin_potential}, we plot the similar figure 
which are scaled with the charm quark mass $\mQ=2.00(5)$ GeV in the scaling factor $1/\mQ^n$.
The largest contribution is still the $\Ocal({1/\mQ^3})$ contribution, while the size of 
higher order contributions becomes comparable to that of the $\Ocal({1/\mQ^3})$ contribution.
Obviously the higher order corrections are much important rather than the leading order correction
of $\Ocal({1/\mQ^2})$ at the charm quark mass.
Nevertheless, the signs of the higher order contributions clearly alternate 
between positive and negative. Remark that the full spin-spin potential is certainly repulsive 
in a whole range of $r$ measured here. 
The higher order contributions of $\Ocal({1/\mQ^4})$ and $\Ocal({1/\mQ^5})$ are 
almost canceled with each other and then the $\Ocal({1/\mQ^3})$ contribution 
approximately represents a whole nature of repulsion of the full spin-spin potential.

These observations may indicate that the $1/\mQ$ expansion is no longer converged 
in the charm quark mass region. In this sense, the new determination of the interquark potential 
{\it at finite quark mass} within the BS amplitude method is a powerful tool for exploring the charmonium system. 
We can compute theoretical inputs for modeling the reliable interquark potential 
from first principles QCD, and then provide new and valuable information to especially 
the spin-dependent potentials including the tensor and spin-orbit potentials in the quark potential models.

Finally, we make a comment on the peculiar behavior found in the $\Ocal({1/\mQ^2})$
contribution near the origin. We first recall that a residual discretization error that 
may not be removed in the RHQ action is of order $\Ocal ((ap)^2(am_Q))$. 
The inverse of lattice spacing for the FI ensembles used here is about 4.2 GeV, which is not quite higher than the bottom quark mass,
rather lower than our three heaviest quark masses ($\mQ = 4.57(7)$, 5.80(7) and 7.71(8) GeV)
in this study.  Therefore, our data set of the interquark potentials in principle suffers from the 
the residual discretization error, which may not be serious in simulations at the charm quark mass. 
In the analysis discussed here, the data obtained at heavier quark masses is obviously
important. Therefore, the final results, which highly relies on the heavy quark mass extrapolation, 
should receive some influence of the residual discretization error, which is not negligible in the
short distance region of $r \alt 1/p \alt a (a\mQ)^{1/2} \sim 0.1~\text{fm}$.
Therefore, in above discussions, we simply disregard the short-range behavior that we should not take seriously.

 %
 %
\section{\label{summary}Summary}
We have proposed the new method to determine the interquark potential
at finite quark mass in lattice QCD.
The $\QQbar$ potential is defined through the equal-time $\QQbar$ Bethe-Salpeter wave function
and also the quark kinetic mass is self-consistently determined on the same footing.
The proper definition of the quark mass is essential for the application of the BS amplitude
method to the $\QQbar$ system. 
The spin-independent and dependent parts of interquark potential together with the quark kinetic mass 
can be calculated with a single set of four-point correlation functions.

We have demonstrated the feasibility of our new proposal by using quenched lattice QCD simulations.
In order to study several systematic uncertainties on the interquark potential,
our simulations were performed on several gauge ensembles generated in the quenched approximation 
at three different lattice spacings, $a \approx  0.093, 0.068$ and  $0.047$~fm,
and two different physical volumes, $La\approx 2.2$ and $3.0 $~fm.
The heavy quark propagators were computed using the RHQ action
with the coefficients determined by one-loop perturbative calculations.
The hopping parameter was chosen to reproduce the experimental spin-averaged mass 
of the $1S$ charmonium states.

In the BS amplitude method, there is a room for optimizing the differential operator since
the discrete Laplacian operator is itself built in the definition of the interquark potential.
Through the simulations carried out at three different lattice spacings, we first conclude 
that the discrete Laplacian operator in the discrete polar coordinates is 
more suitable than the naive one defined in the Cartesian coordinates to reduce the 
discretization artifacts on the short-range behavior of the interquark potential.
The resultant spin-independent central potential in quenched lattice QCD exhibits 
the linearly rising potential at large distances and the Coulomb-like potential at short distances.
All results calculated at three different lattice spacings nicely collapse on a single curve.
In this sense,  the rotational symmetry 
is effectively recovered in the spin-independent central potential calculated in the BS amplitude
method. We also confirm, through simulations on two different physical volumes, that the
finite volume effect on the interquark potential is negligible if the BS wave function
safely fits into the lattice volume used for the simulation.

We have additionally examined the quark mass dependence of the interquark potential 
over a wide mass range from the charm to beyond the bottom toward the infinitely heavy quark limit,
using the finest lattice spacing ensembles~(the inverse of lattice spacing is $1/a_{48} \approx 4.2$~GeV).
We then demonstrated that the spin-independent central potential in the $\mQ \to \infty$ limit 
is connected to the static interquark obtained from Wilson-loops and Polyakov lines, and 
find that the $\Ocal(1/\mQ^2)$ correction should be non-negligible on the short-range part 
of the spin-independent central potential at around the charm quark mass. 

The spin-spin potential at finite quark mass in quenched lattice QCD
provides not pointlike, but finite-range repulsive interaction.
The spin-spin potential determined in the new method potentially
accounts for all orders of $1/\mQ$ corrections, and also shows the
qualitative difference from the slightly attractive spin-spin potential measured 
at $\Ocal(1/\mQ^2)$ in pNRQCD.
The repulsive feature of the spin-spin interaction is phenomenologically 
required by the observed mass-ordering found in hyperfine multiplets.
The issue on the spin-spin potential determined in the $1/\mQ$ expansion 
approach may be resolved by what we found in a detailed study of quark mass dependence
on the spin-spin potential calculated by the BS amplitude method.

We read off from our spin-spin potential, which may encode all orders of the $1/\mQ$ expansion,
that the corresponding $\Ocal(1/\mQ^2)$ correction exhibits the slight attraction
and then barely agrees with the Wilson-loop results. Furthermore, the most important contribution to
the spin-spin potential should be the $\Ocal(1/\mQ^3)$ correction, which is 
responsible for the repulsive feature of the total spin-spin potential, 
rather than the $\Ocal(1/\mQ^2)$ correction even at the bottom quark mass.

We finally conclude that both the spin-independent central and spin-spin potentials
calculated at finite quark mass in the BS amplitude method can reproduce known results 
calculated within the Wilson loop formalism in the infinitely heavy quark limit.
Apparently the new method proposed in this paper has the advantage of determining
the proper $\QQbar$ potential in not only the bottom sectors, but also the charm sector.

From the viewpoint of phenomenology, greater knowledge of the $r$-dependence of the spin-dependent 
potentials paves way for making more accurate theoretical predictions about the higher-mass quarkonium states.
Indeed, the $r$-dependence of the spin-spin potential calculated from first principles of QCD is 
significantly different from a repulsive $\delta$-function potential of the Fermi-Breit interaction, 
which is widely adopted in quark potential models.

In this sense, a full set of the reliable spin-dependent potentials derived from lattice QCD 
can provide new and valuable information to the quark potential models.
We plan to develop our method to determine all spin-dependent potentials including
the tensor and spin-orbit forces. The tensor one is especially required to quantify the size of a mixing 
between $1S$ and $1D$ wave functions, that is assumed to be negligible in the vector
quarkonium states in our current analysis. Such planning is now under way.

 %
 %
\begin{acknowledgments}
 It is a pleasure to acknowledge helpful suggestions with T. Hatsuda, and to thank H. Iida and Y. Ikeda 
 for fruitful discussions. S.S. also thanks to Y. Sumino for discussion 
 on the next-to-leading order term of the spin-spin potential in the $1/\mQ$ expansion.
 This work is supported by JSPS Grants-in-Aid for Scientific Research (No.~23540284).
 T.K. is partially supported by JSPS Strategic Young Researcher Overseas Visits Program
  for Accelerating Brain Circulation (No.~R2411).
\end{acknowledgments}

\bibliography{paper} 
\end{document}